\documentclass[%
reprint,
superscriptaddress,
 amsmath,amssymb,
 aps,
floatfix,
]{revtex4-2}

\usepackage{graphicx}% Include figure files
\usepackage{dcolumn}% Align table columns on decimal point
\usepackage{bm}% bold math
\usepackage{xcolor} % Jon added on 4/5 for adding notes.

\newcommand{\Section}[1]{{\emph{#1}}} %\textbf

\begin{document}

\preprint{APS/123-QED}

%\title{Imaging Direct MOT-Loading and Bright $\Lambda$-Enhanced Grey Molasses Imaging of $^6$Li in Optical Tweezers}
\title{Imaging a $^6$Li Atom  In An Optical Tweezer 2000 Times with  $\Lambda$-Enhanced Gray Molasses}

\newcommand{\affa}{\affiliation{Department of Chemistry, Purdue University, West Lafayette, Indiana, 47907, USA}}
\newcommand{\affb}{\affiliation{Department of Physics and Astronomy, Purdue University, West Lafayette, Indiana, 47907, USA}}

\author{Karl N. Blodgett}  \affa
\author{David Peana} \affa
\author{Saumitra Phatak}  \affb
\author{Lane M. Terry} \affa  %or use current address
\author{Maria Paula Montes}  \affb
\author{Jonathan Hood} \email{hoodjd@purdue.edu} \affa \affb

\date{\today}

\newcommand\loadingrate{0.68(3)}
\newcommand\fidelity{0.9997(2)}
\newcommand\survival{0.99950(2)}
\newcommand\HoldLifetime{5.13(7)}  %5.126
\newcommand\GMLifetime{30.0(8)}
\newcommand\LossRate{0.00050(1)}
%\the\textwidth  %510 pt  % \the\columnwidth   246 pt % width to make figures

\begin{abstract}
We have imaged lithium-6 thousands of times in an optical tweezer using $\Lambda$-enhanced gray molasses cooling light. Despite being the lightest alkali, with a recoil temperature of 3.5 $\mu$K, we achieve an imaging survival of \survival{}, which sets the new benchmark for low-loss imaging of neutral atoms in optical tweezers.  Lithium is loaded directly from a MOT into a tweezer with an enhanced loading rate of 0.7. We cool the atom to 70 $\mu$K and present a new cooling model that accurately predicts steady-state temperature and scattering rate in the tweezer. These results pave the way for ground state preparation of lithium en route to the assembly of the LiCs molecule in its ground state.  
\end{abstract}

\maketitle

Optical tweezers are a dynamic platform for manipulating ultracold atoms and molecules in a variety of applications, including few-body physics with neutral atoms \cite{kaufman2014two, spar2022realization}, long-range interactions with arrays of Rydberg atoms \cite{levine2018high,ma2022universal}, and atom-by-atom molecular assembly and dipolar interaction \cite{liu2019molecular, spence2022preparation, wang2019preparation, zhang2020forming,holland2022demand, bao2022dipolar}.  The fluorescence of a single atom can be imaged with high fidelity, allowing for single site control and rearrangement of a partially filled array into arbitrary, low entropy configurations~\cite{endres2016atom,barredo2018synthetic}.  

Imaging survival probability affects gate fidelity readout with Rydberg atoms, and is a limiting factor in molecular assembly and the production of defect-free arrays \cite{endres2016atom,liu2018building}. Tweezer-trapped alkali atoms have been conventionally imaged with polarization gradient (PG) cooling light \cite{kaufman2014two, yu2018motional} with survival probability per image  of  approximately 0.98. Recently, $\Lambda$-enhanced gray molasses ($\Lambda$-GM) cooling has been used in tandem with resonant scattering light to image $^{39}$K with a survival rate exceeding 0.99 \cite{huang2022gray}. The narrow optical transition lines of alkaline-earth atoms (AEAs) make them suitable for new cooling techniques, with which the survival rate has been pushed to 0.99932(8), the current benchmark for low-loss neutral atom imaging \cite{covey20192000}. 

\begin{figure*}[t!]
    \centering
    \includegraphics[width=\textwidth]{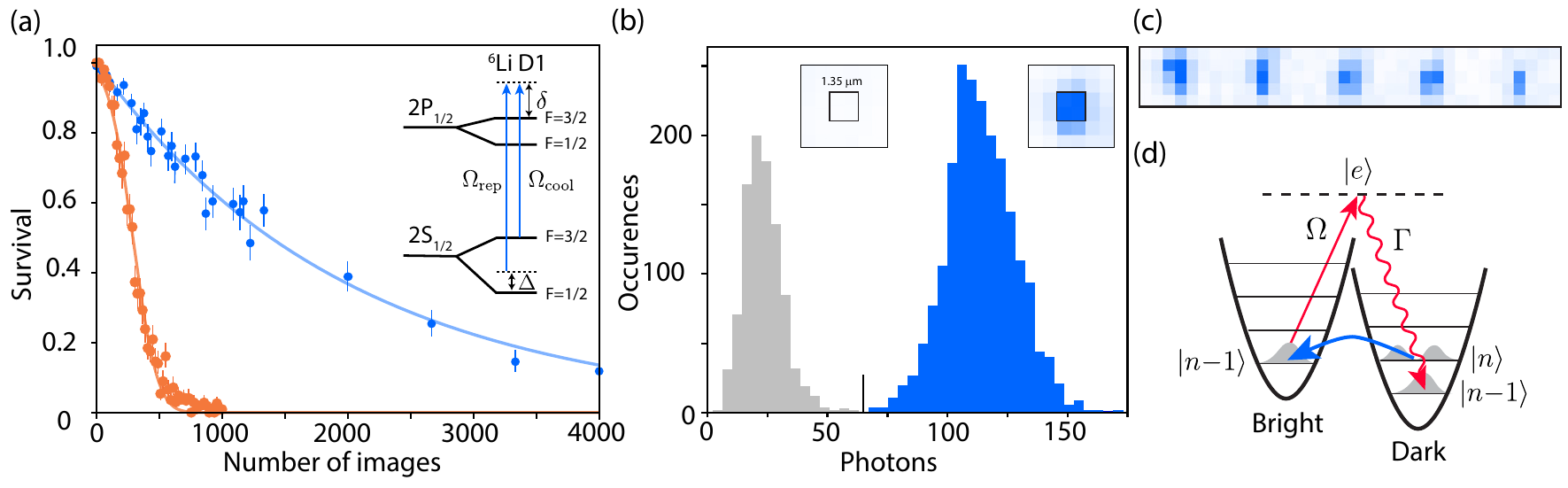}
    \caption{\label{fig1} \textbf{Low-loss $\Lambda$-GM imaging of $^6$Li.} (a) Atom survival in an optical tweezer with (blue) and without (orange) active cooling light vs. consecutive images.  The data with continuous GM cooling has a $1/e$ decay time of \GMLifetime{} s, corresponding to 2000 images.  The atom without active cooling light applied is heated due to off-resonant scattering of the tweezer light and has a $1/e$ time of \HoldLifetime{} s.  The inset is the D1 level structure with $\Lambda$-GM parameters labelled.
    (b) Histogram of fluorescence from a single $^6$Li atom trapped in a 1064 nm optical tweezer. This histogram represents an averaged image (3000 shots) of a single region-of-interest with a signal threshold of 65 photon counts and a loading rate of \loadingrate{}. Fitting a Gaussian distribution to both peaks and integrating the 0 atom (1 atom) fraction above (below) the signal threshold gives an imaging fidelity of \fidelity{}. The insets show the cropped region-of-interest of an image of 0 (left) or 1 (right) atom.   
    (c) Single-shot image of five occupied adjacent tweezer sites.
    (d) Simplified picture of $\Lambda$-GM in a tweezer.  A dark state $|D, n \rangle$ couples to the bright state $|B, n-1 \rangle$ with one less motional quanta.  Spontaneous emission then pumps the population to the dark state $|D, n-1 \rangle$ when in the Lamb-Dicke regime.}
\end{figure*}

In this letter, we demonstrate imaging of lithium in an optical tweezer using $\Lambda$-enhanced GM light with a survival probability of \survival{} (Fig.~\ref{fig1}(a)). This represents an order of magnitude improvement over survival probability in other alkali atoms, and even surpasses the best results with narrow-line AEAs. The 0 and 1 atom cases are distinguished with a fidelity of \fidelity{}, which comes as an order of magnitude improvement over existing lithium single atom imaging techniques \cite{parsons2016site,brown2017spin,bergschneider2018spin,holten2021observation}. 

We develop a numerical model based on the master equation that accurately predicts the steady-state temperature and scattering rate in the tweezer. As shown in Fig.~\ref{fig1}(d), motional energy is removed by Raman coupling from a dark state to a bright state with one less motional quanta, followed by optical pumping back to a dark state.  Our model suggests that this scheme may be used to cool trapped lithium atoms along one motional axis at a time, similar to existing motional cooling techniques but with a marked reduction in experimental complexity. This would lay the groundwork for the formation of the LiCs molecule in its ground state. With the largest dipole moment (5.5 Debye) of any bi-alkali molecule, LiCs is a promising candidate for molecule-based quantum gates.

\begin{figure}[]
    \centering
    \includegraphics[width=.35\textwidth]{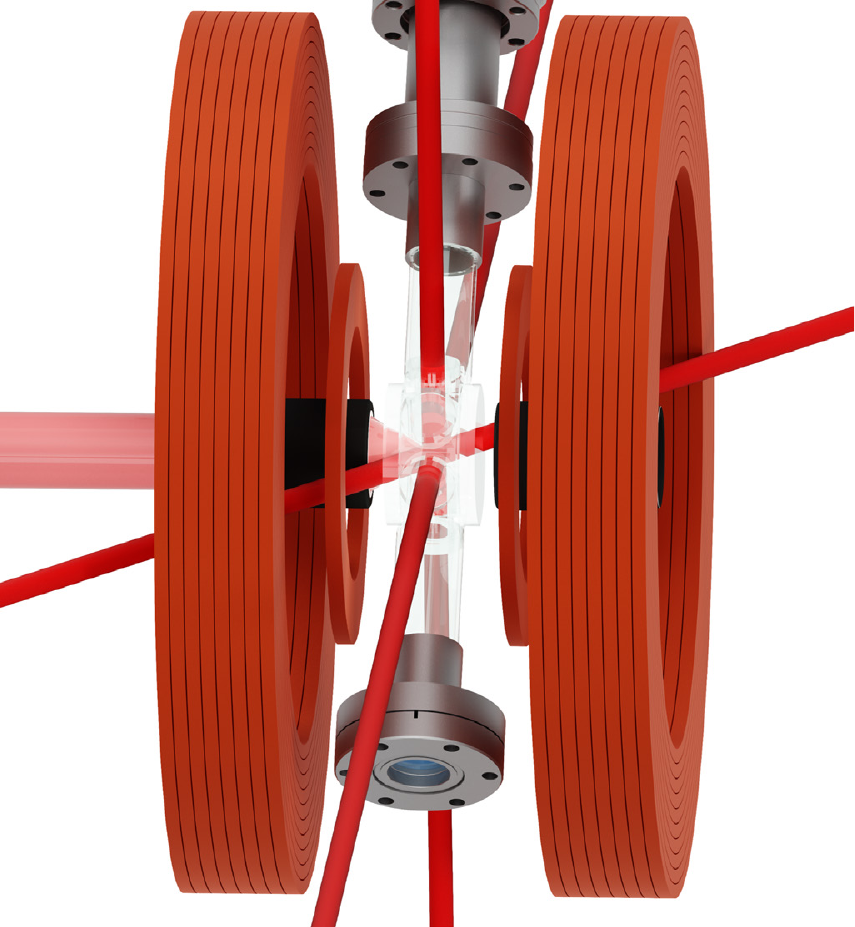}
    \caption{\label{fig2} \textbf{Single-atom imaging.} The 1064 nm tweezer light (light red) is focused down to $\sim$1 $\mu$m at the center of the double-sided glass octagon cell. Three retroreflected laser beams containing $\Lambda$-GM light (dark red) are sent to the tweezer-trapped atom(s). Fluorescent photons are collected back through the same objective and imaged onto an EMCCD camera. MOT and mini-MOT coils are shown, shim coils are omitted for clarity.}
\end{figure}

%\Section{Results:}
The experiment begins with using the D2 line to load a 3-D magneto-optical trap (MOT) of approximately 10$^5$ $^6$Li atoms in 100 ms in the science region of our vacuum chamber (Fig.~\ref{fig2}). Since loading optical tweezers requires relatively few atoms, we use the radial decay of the quadrupole magnetic field from the MOT coils, rather than a conventional tapered-coil design, for our Zeeman field. 

\begin{figure}[]
    \centering
    \includegraphics[width=.99\columnwidth]{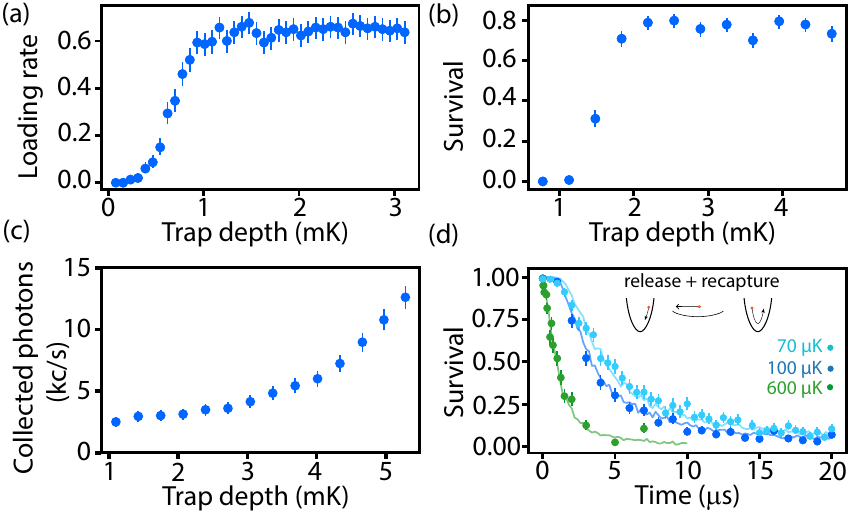}
    \caption{\label{fig3} \textbf{Properties of $^6$Li in a tweezer.} (a) Atom loading vs trap depth (mK). For imaging, trap depth is ramped to a constant value of 3.4 mK. (b) Survival vs trap depth after 533 consecutive image cycles. (c) Photons collected vs trap depth during 15 ms image. In (b) and (c), $\Lambda$-GM parameters are not changed as a function of trap depth. (d) Release-and-recapture survival probability under different cooling scenarios. Light blue: optimized cooling for low temperature. Dark blue: optimized cooling for survival. Green: hold in tweezer for 1 s (no cooling light applied). Solid lines are Monte Carlo simulations used to extract a temperature.}
\end{figure}

To initiate tweezer loading, the 3D MOT is overlapped with a 1064 nm tweezer beam that is generated from a low-noise, high-power fiber laser. Magnetic fields are switched off and 3 ms of $\Lambda$-GM is applied to the D1 line of the atomic cloud, cooling the atoms to sub-doppler temperatures of 100 $\mu$K. The optimum free-space cooling conditions are a two-photon detuning $\Delta$ = 0 MHz, single-photon detuning $\delta=30$~MHz, $I_{\text{rep}}/I_{\text{cool}} = 1/10$, and $I/I_{\text{sat}} = 15.0$. The repumper light is generated as a coherent sideband of the cooler light with an electro-optic modulator. We observe the onset of tweezer loading at a trap depth of $U$/$k_B$ $\approx$ 0.3 mK, and saturation to an enhanced loading rate of 0.7 at 1.2 mK (Fig.~\ref{fig3}(b)). 

The tweezer light is split into an array of individual trap sites by an acousto-optic deflector (AOD) before being focused down to the center of our octagon glass cell by a high numerical aperture objective (NA $=0.55$). Radial trap frequency measurements in combination with the measured trap depth yield orthogonal radial beam waists of 0.98 and 1.15 $\mu$m. 

To image $^6$Li in a tweezer, we switch on D1 $\Lambda$-GM light for 15 ms and send it through the same path as the three D2 MOT beams, two of which are orthogonal, while one traverses a 30$^{\circ}$ angle with respect to the tweezer propagation axis (Fig.~\ref{fig2}). Scattered 671 nm light is collected through the same objective that the tweezer light is focused through and imaged onto an EMCCD camera. 

Fig.~\ref{fig1}(b) presents a histogram of collected photons under typical imaging conditions for a single 3.4 mK deep tweezer site. We are able to discriminate between instances of 0 and 1 atom in the tweezer with very high fidelity (\fidelity{}), as indicated by fitting the bimodal distribution with two Gaussian distributions. 
Parity projection of the initial Poisson-distributed number of atoms loaded into the trap from several to either 0 or 1 is achieved via an 8 ms light-assisted collision (LAC) pulse. Inspired by the implementation of blue-detuned LAC light in loading rate enhancement beyond the expected stochastic value of 0.5 \cite{brown2019gray,grunzweig2010near,lester2015rapid,huang2022gray}, we use $\Lambda$-GM for our parity projection pulse to achieve loading rates of up to 0.7.

%\subsection{\label{sec:survival}Survival Dynamics}
%\Section{Survival:}
We optimized $\Lambda$-GM in the tweezer with respect to either atom survival or temperature.  For a 3.4 mK tweezer, we find optimum survival conditions by setting $\Delta$ = 0 MHz, $\delta$ = 20 MHz, and $I/I_\text{sat}$ = 9.3. Despite the dark states of the $\Lambda$ system, we collect photons at our camera at a rate of 6 kc/s resulting in a high imaging fidelity in only 15 ms.  %85/15ms counts/s detection rate, or 5700 1/s.   With rho = 0.006, then scattering rate of 220,000 1/s,  meaning we are collecting 2.6% of photons. 
This light proved to be more than sufficient for atom imaging \cite{cheuk2018lambda}. Under survival-optimized conditions, we measure a temperature of 100 $\mu$K using the release-and-recapture method~\cite{tuchendler2008energy} (Fig.~\ref{fig3}(d), dark blue).

We measure the atom survival under constant application of $\Lambda$-GM light. Conditioned on whether an atom was present in the first image, the occupancy of an atom in the second image is counted towards the survival statistics for a given hold time. We plot the survival probability of the atom as a function of imaging cycles in Fig.~\ref{fig1}(a). The blue data shows the imaging survival versus the number of 15 ms imaging cycles and gives a $1/e$ lifetime of \GMLifetime{} s, corresponding to 2000 images, or a loss rate per image of \LossRate{}.

We also measure the atom survival time without any cooling light to obtain a ${1/e}$ lifetime of \HoldLifetime{} seconds. Care was taken to ensure that atom loss is not due to the lithium atomic beam or leaked light. This increased loss rate is consistent with heating from the 1064 nm trap laser with an inelastic scattering rate of $\sim$10 Hz in combination with lithium's large recoil temperature of 3.5~$\mu$K. The loss curve fits an error function rather than an exponential, indicating heating. This is evident by measuring an increased temperature of 600 $\mu$K after 1~s of simply holding the atom in the tweezer (Fig.~\ref{fig3}(d), green).

The lowest temperature reached using $\Lambda$-GM in the tweezer is 70~$\mu$K, where relative to survival conditions, $\delta$ increased by $\sim$ $\Gamma$, and the cooling and repumper power decrease by a factor of two. Interestingly, the atom imaging $1/e$ lifetime under these conditions is nearly identical to that of holding the atom, indicating that the temperature distribution has a larger population at higher $n$.  

%\Section{Analysis:}
The imaging survival probability of $^6$Li with  $\Lambda$-GM cooling is more than an order of magnitude better than previous alkali results in a tweezer. The high fluorescence rate also allows for even higher imaging fidelity in only 15 ms. To understand this improvement, we simulate $\Lambda$-GM cooling for $^6$Li in a harmonic trap by solving the steady-state solution of the full master equation for $^6$Li with both the cooling  and repumper beams in $\sigma_+$-$\sigma_-$ configuration (detailed further in the SM~\cite{supp}). 
%Polarization gradient~\cite{dalibard1989laser} and 
GM and $\Lambda$-GM has been described by free-space models with classical atoms moving through polarization gradients that vary on the scale of the laser wavelength~\cite{salomon2014gray}. The motion of the atom couples the dark states to the bright states, where the atom travels up a potential energy hill before being pumped back to a dark state in a Sisyphus style cooling. 
This picture works well in free space, but breaks down when the atom is constrained to a length scale much smaller than the wavelength, i.e., when the trapped atom is in the Lamb-Dicke (LD) regime. This discrepancy motivates the development of a new picture of $\Lambda$-GM in a tweezer.

%Cooling is achieved in gray molasses by a having a bright state with a stark shift modulated within a wavelength, a lower energy dark state, and a coupling between them due to motion.  $\Lambda$-enhancement occurs when coherent lasers are resonant to the two hyperfine ground states, producing extra dark states in the form of the the bright symmetric and dark anti-symmetric combinations of the three-level system.

%Inspired by the photon sideband cooling formalism~\cite{diedrich1989laser}  \Jon{check ref},
The LD parameter is defined as the ratio of the recoil energy to the trapping frequency energy, $\eta_{LD} = (E_\text{recoil}/ \hbar \omega )^{1/2} \!=\! k x_0  $,  where $x_0 \!=\! \sqrt{\hbar/2m\omega}$ is the ground state harmonic length scale, $k$ is the wave-vector of the cooling light, and $\omega$ is the trapping frequency.  The atom-field interaction of a plane wave of light in the rotating frame is  $H_{AF} = \Omega \, \hat{\sigma}^\dag \, e^{ i k \hat{x} } + h.c. $, where $\hat{\sigma}$ is the atom lowering operator, and $\hat{x}$ is the atom position operator. 
In the LD regime, where $\eta_{LD}\ll$ 1, the interaction can be expanded as $ H_{AF} \approx  \Omega \, \hat{\sigma} +  \eta_\text{LD} \, \Omega \, \hat{\sigma}^\dag  (\hat{a} + \hat{a}^\dag) + h.c. $, where the zeroth-order term couples states with the same $n$, and the first-order terms couples states with different $n$, resulting in heating or cooling.

\begin{figure}[t] 
    \centering
    \includegraphics[width=.99\columnwidth]{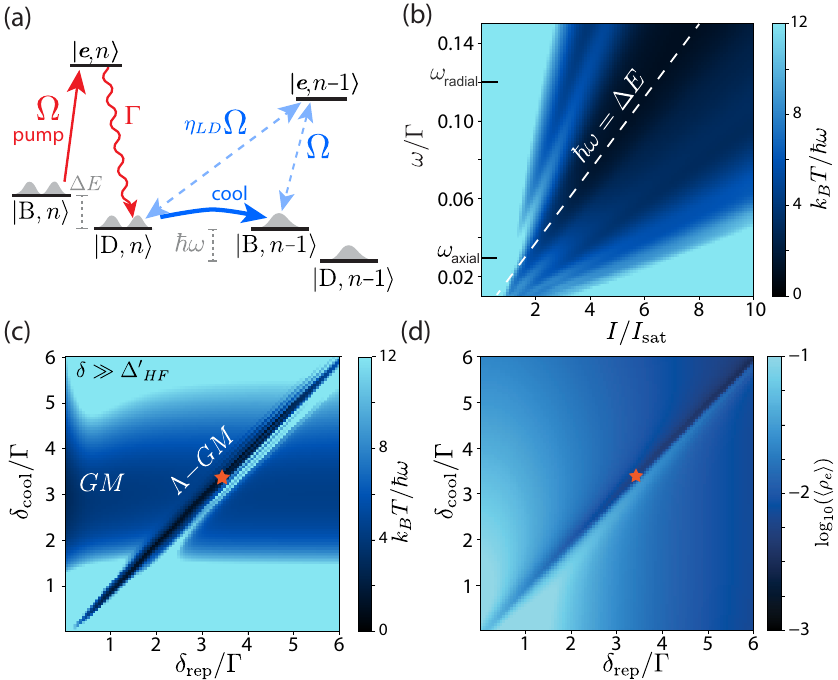}
    \caption{\label{fig4} \textbf{Simulation of $\Lambda$-GM in a tweezer.} (a) $\Lambda$-GM in a harmonic potential. In step 1, spontaneous emission pumps the bright state $|B, n \rangle$ to the dark state $|D,n \rangle$ with the same motional level.  In step 2, the dark state is transferred to the bright state with a net loss of one motional quanta due to the three-level system formed by the $\Omega$ and $ \eta_{LD}\Omega$ terms. (b) Simulation of steady-state temperature as a function of cooling power and trapping frequency for $\delta_1= \delta_2= $ 20 MHz.  The dashed line is where the bright state shift equals the trapping frequency. (c) Temperature versus the cooling and repumper detunings. $\Lambda$-GM is seen on the diagonal where $\Delta=0$ MHz. Normal GM is seen away from that condition.   (d) Plot of population of steady-state in the excited state. The experimental optimum is 0.006 excited, giving a scattering rate consistent with experiment.}
\end{figure}

Cooling can be understood by the simplified picture shown in Fig.~\ref{fig4}(a).  
When $\Omega_\text{cool}$ and $\Omega_\text{rep}$  couple the $F=3/2$ and $F=1/2$ ground states resonantly ($\Delta=0$ MHz) with a blue single-photon detuning $\delta$, we can go into a dressed basis of the zeroth-order LD expansion to get bright $|B,n \rangle$ and dark $|D,n \rangle$ states, where $n$ is the harmonic level.
 The bright states are AC Stark shifted upwards by $\Delta E$ due to the blue-detuned light and are coupled to the excited state $|e, n \rangle$ by $\Omega$, while the dark state is not AC Stark shifted or coupled to the excited state. %by the effective Raman coupling $\Omega_R = \Omega_\text{cool} \Omega_\text{rep}/\delta $.  
 The neighboring motional state $n-1$ is also drawn and shifted down by the trapping frequency $\hbar \omega$.   

The cooling is described in two steps: First (Fig.~\ref{fig4}(a), red), spontaneous emission pumps the population from the bright state $|B, n \rangle$ to the dark state $|D, n \rangle$ without changing the motional level; this is followed by a second step (Fig.~\ref{fig4}(a), blue), in which the population is transferred from the dark state  $|D, n\rangle$  to the bright state $|B, n -1 \rangle$, with a net loss of one motional quanta. Bright/dark state coupling occurs due to the interaction terms  $\Omega \hat{\sigma}^\dag$  and $\eta_{LD} \Omega \hat{\sigma}^\dag \hat{a}  $, which couple each state to a common excited state, $|e, n \rangle$, forming a three-level system in which population is transferred via a coherent Raman transition.  The transfer is most efficient when the bright state Stark shift $\Delta E$ is similar to the motional trapping frequency $\hbar \omega$, where the Raman coupling from $|D,n \rangle $ to $|B, n-1 \rangle$ is resonant and strongest.  This condition, $\hbar \omega = \Delta E $, is drawn in the simulation Fig.~\ref{fig4}(b) as a white dashed line and shows good agreement with optimal cooling. 

%In Fig.~\ref{fig4}(b), we simulate cooling for $^6$Li by solving the steady state master equation, the details of which are described in the SM~\cite{supp}.  The simulation includes the two hyperfine levels in the ground and D1 state.  We do not assume the LD regime for the interaction, and includes cross-coupling of the two frequencies.  We use $\sigma^+$-$\sigma^-$ configuration in 1D, which gives a rotating linear polarization.  The atom will therefore see linear polarization and experience to zero'th order in LD regime a tensor shift, with a different shift for the $\Delta E_{m\pm1/2}$ and  $\Delta E_{m\pm3/2}$ states (see inset). 
%The electric field in the LD expansion that has a is orthgonal because of the counter propagating beam, resulting in a mixing of the bright adn dark states when changing $n$. 

Fig.~\ref{fig4}(b) shows a simulation of the steady-state temperature  as a function of cooling power and the trapping frequency $\omega$.  The simulation parameters match the experimental settings described above, with $\Omega_\text{rep}\! = \! \sqrt{0.1} \, \Omega_\text{cool}$, $\delta = 20 $ MHz, and $\Delta=0$ MHz.  The temperature is obtained by fitting a Boltzmann distribution to the steady-state motional populations~\cite{supp}. Similar to the experiment, we  find in the simulation that the temperature is lowest for $\Omega_\text{rep}  \ll \Omega_\text{cool}$ because it results in less scattering in the dark states, which, in the limit of low repumper intensity, are just the lower hyperfine states $F=1/2$.  The axial and radial trapping frequencies are marked on the y-axis.  The simulated temperatures are 41 $\mu$K for radial and 82 $\mu$K for axial, where the summed averaged energies results in a temperature of 92 $\mu$K, in good agreement with the measured release-and-recapture temperature. These temperatures correspond to 43$\%$ and 11$\%$ motional ground state population for the radial and axial dimensions, respectively. 

% hw/kB = 33.6 uK for 700 kHz.  
% from 2d simulations in paper,  T = 1.2 (hw/kBT)^-1,  or  41 uK for radial,  <n> = 3.0
% for axial 12,  or  82 uK.  <n> = 5.0
% 0.006 excited for both.  
% (2*(<nradial> +1/2)*34uK + (<naxial>+1/2)*34/5 uK) = 3 kBT,  or T = 92 uK. 
% for optimized radial and axial separately, 0.8 hbar omega/kT =27 uK for radial or 57% ground state, and 3.7 hbar omega/kT axial, corresponding to 23% ground state.  , or 7% 3D ground state.

%$ \langle n_\text{radial}  \rangle = XX$   and $ \langle n_\text{axial}  \rangle$ \Jon{XX}, giving a simulated temperature of \Jon{XX} $\mu$, in good agreement with the measured temperature of 100 $\mu$K. \Jon{check these}
Crucially, we see that the axial and radial cooling have different optimal conditions. Experimentally, we have optimized the combined temperature, but the simulation suggests we could cool further by pulsing the radial and axial cooling separately with different powers and detunings.  

In Fig.~\ref{fig4}(c), we plot the radial temperature as a function of the two single-photon detunings, with  $I/I_\text{sat} = 6$.   The star marks marks the experimentally determined survival optimum.   The simulation shows gray molasses (GM) cooling centered around $\delta_\text{cool} \approx 3 \Gamma$, where it decays as the detuning increases beyond the excited-state hyperfine splitting of 26 MHz where the tensor shift is canceled out by the coupling to both excited hyperfine states. Notably, GM is largely insensitive to the detuning of the low-intensity repumper laser. $\Lambda$-enhancement is found along the diagonal where $\Delta \approx 0 $ MHz, where a corresponding factor of two decrease in temperature is predicted relative to standard GM.  

Fig.~\ref{fig4}(d) plots the excited state population of the steady-state solution from Fig.~\ref{fig4}(c).  The simulation predicts a total scattering rate of 220 kc/s, in good agreement with our collected photon rate of 6 kc/s and a collection efficiency predicted to be  $\sim$3$\%$.  We also find an explanation for why lithium $\Lambda$-GM in a tweezer is relatively bright - the  small excited hyperfine splitting  ruins the three-level system model and prevents the dark state from becoming greater than 99.5$\%$ dark under any conditions. 

%\Section{Conclusion:}
In conclusion, we have loaded lithium-6 directly from a MOT into optical tweezers with an enhanced loading rate in 100 ms. Individual lithium atoms have been imaged thousands of times with $\Lambda$-enhanced gray molasses light, where, despite its light mass, high recoil energy, and lack of accessible narrow lines, we obtain a survival probability of \survival{}, which sets the benchmark for low-loss neutral atom imaging in a tweezer. Additionally, an imaging fidelity of \fidelity{} improves upon existing lithium single atom imaging by an order of magnitude. 

We have simultaneously developed a model that describes $\Lambda$-GM in the tweezer, which accurately predicts scattering rate and steady-state temperature. Our model elucidates new strategies for motional cooling in the tweezer, which could pave the way towards ground state preparation of lithium as a prelude to LiCs molecular assembly.

\Section{Acknowledgements:}
We thank Ben Cerjan, Phil Wyss, Mark Carlsen, and the Jonathan Amy Facility for Chemical Instrumentation (JAFCI) for valuable contributions to the instrumentation. This work was supported by the NSF Career Award (Award No. 0543784). 

\bibliography{bibliography}

%apsrev4-2.bst 2019-01-14 (MD) hand-edited version of apsrev4-1.bst
%Control: key (0)
%Control: author (72) initials jnrlst
%Control: editor formatted (1) identically to author
%Control: production of article title (-1) disabled
%Control: page (0) single
%Control: year (1) truncated
%Control: production of eprint (0) enabled
\providecommand{\noopsort}[1]{}\providecommand{\singleletter}[1]{#1}%
\begin{thebibliography}{27}%
\makeatletter
\providecommand \@ifxundefined [1]{%
 \@ifx{#1\undefined}
}%
\providecommand \@ifnum [1]{%
 \ifnum #1\expandafter \@firstoftwo
 \else \expandafter \@secondoftwo
 \fi
}%
\providecommand \@ifx [1]{%
 \ifx #1\expandafter \@firstoftwo
 \else \expandafter \@secondoftwo
 \fi
}%
\providecommand \natexlab [1]{#1}%
\providecommand \enquote  [1]{``#1''}%
\providecommand \bibnamefont  [1]{#1}%
\providecommand \bibfnamefont [1]{#1}%
\providecommand \citenamefont [1]{#1}%
\providecommand \href@noop [0]{\@secondoftwo}%
\providecommand \href [0]{\begingroup \@sanitize@url \@href}%
\providecommand \@href[1]{\@@startlink{#1}\@@href}%
\providecommand \@@href[1]{\endgroup#1\@@endlink}%
\providecommand \@sanitize@url [0]{\catcode `\\12\catcode `\$12\catcode
  `\&12\catcode `\#12\catcode `\^12\catcode `\_12\catcode `\%12\relax}%
\providecommand \@@startlink[1]{}%
\providecommand \@@endlink[0]{}%
\providecommand \url  [0]{\begingroup\@sanitize@url \@url }%
\providecommand \@url [1]{\endgroup\@href {#1}{\urlprefix }}%
\providecommand \urlprefix  [0]{URL }%
\providecommand \Eprint [0]{\href }%
\providecommand \doibase [0]{https://doi.org/}%
\providecommand \selectlanguage [0]{\@gobble}%
\providecommand \bibinfo  [0]{\@secondoftwo}%
\providecommand \bibfield  [0]{\@secondoftwo}%
\providecommand \translation [1]{[#1]}%
\providecommand \BibitemOpen [0]{}%
\providecommand \bibitemStop [0]{}%
\providecommand \bibitemNoStop [0]{.\EOS\space}%
\providecommand \EOS [0]{\spacefactor3000\relax}%
\providecommand \BibitemShut  [1]{\csname bibitem#1\endcsname}%
\let\auto@bib@innerbib\@empty
%</preamble>
\bibitem [{\citenamefont {Kaufman}\ \emph {et~al.}(2014)\citenamefont
  {Kaufman}, \citenamefont {Lester}, \citenamefont {Reynolds}, \citenamefont
  {Wall}, \citenamefont {Foss-Feig}, \citenamefont {Hazzard}, \citenamefont
  {Rey},\ and\ \citenamefont {Regal}}]{kaufman2014two}%
  \BibitemOpen
  \bibfield  {author} {\bibinfo {author} {\bibfnamefont {A.}~\bibnamefont
  {Kaufman}}, \bibinfo {author} {\bibfnamefont {B.}~\bibnamefont {Lester}},
  \bibinfo {author} {\bibfnamefont {C.}~\bibnamefont {Reynolds}}, \bibinfo
  {author} {\bibfnamefont {M.}~\bibnamefont {Wall}}, \bibinfo {author}
  {\bibfnamefont {M.}~\bibnamefont {Foss-Feig}}, \bibinfo {author}
  {\bibfnamefont {K.}~\bibnamefont {Hazzard}}, \bibinfo {author} {\bibfnamefont
  {A.}~\bibnamefont {Rey}},\ and\ \bibinfo {author} {\bibfnamefont
  {C.}~\bibnamefont {Regal}},\ }\href@noop {} {\bibfield  {journal} {\bibinfo
  {journal} {Science}\ }\textbf {\bibinfo {volume} {345}},\ \bibinfo {pages}
  {306} (\bibinfo {year} {2014})}\BibitemShut {NoStop}%
\bibitem [{\citenamefont {Spar}\ \emph {et~al.}(2022)\citenamefont {Spar},
  \citenamefont {Guardado-Sanchez}, \citenamefont {Chi}, \citenamefont {Yan},\
  and\ \citenamefont {Bakr}}]{spar2022realization}%
  \BibitemOpen
  \bibfield  {author} {\bibinfo {author} {\bibfnamefont {B.~M.}\ \bibnamefont
  {Spar}}, \bibinfo {author} {\bibfnamefont {E.}~\bibnamefont
  {Guardado-Sanchez}}, \bibinfo {author} {\bibfnamefont {S.}~\bibnamefont
  {Chi}}, \bibinfo {author} {\bibfnamefont {Z.~Z.}\ \bibnamefont {Yan}},\ and\
  \bibinfo {author} {\bibfnamefont {W.~S.}\ \bibnamefont {Bakr}},\ }\href@noop
  {} {\bibfield  {journal} {\bibinfo  {journal} {Physical review letters}\
  }\textbf {\bibinfo {volume} {128}},\ \bibinfo {pages} {223202} (\bibinfo
  {year} {2022})}\BibitemShut {NoStop}%
\bibitem [{\citenamefont {Levine}\ \emph {et~al.}(2018)\citenamefont {Levine},
  \citenamefont {Keesling}, \citenamefont {Omran}, \citenamefont {Bernien},
  \citenamefont {Schwartz}, \citenamefont {Zibrov}, \citenamefont {Endres},
  \citenamefont {Greiner}, \citenamefont {Vuleti{\'c}},\ and\ \citenamefont
  {Lukin}}]{levine2018high}%
  \BibitemOpen
  \bibfield  {author} {\bibinfo {author} {\bibfnamefont {H.}~\bibnamefont
  {Levine}}, \bibinfo {author} {\bibfnamefont {A.}~\bibnamefont {Keesling}},
  \bibinfo {author} {\bibfnamefont {A.}~\bibnamefont {Omran}}, \bibinfo
  {author} {\bibfnamefont {H.}~\bibnamefont {Bernien}}, \bibinfo {author}
  {\bibfnamefont {S.}~\bibnamefont {Schwartz}}, \bibinfo {author}
  {\bibfnamefont {A.~S.}\ \bibnamefont {Zibrov}}, \bibinfo {author}
  {\bibfnamefont {M.}~\bibnamefont {Endres}}, \bibinfo {author} {\bibfnamefont
  {M.}~\bibnamefont {Greiner}}, \bibinfo {author} {\bibfnamefont
  {V.}~\bibnamefont {Vuleti{\'c}}},\ and\ \bibinfo {author} {\bibfnamefont
  {M.~D.}\ \bibnamefont {Lukin}},\ }\href@noop {} {\bibfield  {journal}
  {\bibinfo  {journal} {Physical review letters}\ }\textbf {\bibinfo {volume}
  {121}},\ \bibinfo {pages} {123603} (\bibinfo {year} {2018})}\BibitemShut
  {NoStop}%
\bibitem [{\citenamefont {Ma}\ \emph {et~al.}(2022)\citenamefont {Ma},
  \citenamefont {Burgers}, \citenamefont {Liu}, \citenamefont {Wilson},
  \citenamefont {Zhang},\ and\ \citenamefont {Thompson}}]{ma2022universal}%
  \BibitemOpen
  \bibfield  {author} {\bibinfo {author} {\bibfnamefont {S.}~\bibnamefont
  {Ma}}, \bibinfo {author} {\bibfnamefont {A.~P.}\ \bibnamefont {Burgers}},
  \bibinfo {author} {\bibfnamefont {G.}~\bibnamefont {Liu}}, \bibinfo {author}
  {\bibfnamefont {J.}~\bibnamefont {Wilson}}, \bibinfo {author} {\bibfnamefont
  {B.}~\bibnamefont {Zhang}},\ and\ \bibinfo {author} {\bibfnamefont {J.~D.}\
  \bibnamefont {Thompson}},\ }\href@noop {} {\bibfield  {journal} {\bibinfo
  {journal} {Physical Review X}\ }\textbf {\bibinfo {volume} {12}},\ \bibinfo
  {pages} {021028} (\bibinfo {year} {2022})}\BibitemShut {NoStop}%
\bibitem [{\citenamefont {Liu}\ \emph {et~al.}(2019)\citenamefont {Liu},
  \citenamefont {Hood}, \citenamefont {Yu}, \citenamefont {Zhang},
  \citenamefont {Wang}, \citenamefont {Lin}, \citenamefont {Rosenband},\ and\
  \citenamefont {Ni}}]{liu2019molecular}%
  \BibitemOpen
  \bibfield  {author} {\bibinfo {author} {\bibfnamefont {L.~R.}\ \bibnamefont
  {Liu}}, \bibinfo {author} {\bibfnamefont {J.~D.}\ \bibnamefont {Hood}},
  \bibinfo {author} {\bibfnamefont {Y.}~\bibnamefont {Yu}}, \bibinfo {author}
  {\bibfnamefont {J.~T.}\ \bibnamefont {Zhang}}, \bibinfo {author}
  {\bibfnamefont {K.}~\bibnamefont {Wang}}, \bibinfo {author} {\bibfnamefont
  {Y.-W.}\ \bibnamefont {Lin}}, \bibinfo {author} {\bibfnamefont
  {T.}~\bibnamefont {Rosenband}},\ and\ \bibinfo {author} {\bibfnamefont
  {K.-K.}\ \bibnamefont {Ni}},\ }\href@noop {} {\bibfield  {journal} {\bibinfo
  {journal} {Physical Review X}\ }\textbf {\bibinfo {volume} {9}},\ \bibinfo
  {pages} {021039} (\bibinfo {year} {2019})}\BibitemShut {NoStop}%
\bibitem [{\citenamefont {Spence}\ \emph {et~al.}(2022)\citenamefont {Spence},
  \citenamefont {Brooks}, \citenamefont {Ruttley}, \citenamefont {Guttridge},\
  and\ \citenamefont {Cornish}}]{spence2022preparation}%
  \BibitemOpen
  \bibfield  {author} {\bibinfo {author} {\bibfnamefont {S.}~\bibnamefont
  {Spence}}, \bibinfo {author} {\bibfnamefont {R.}~\bibnamefont {Brooks}},
  \bibinfo {author} {\bibfnamefont {D.}~\bibnamefont {Ruttley}}, \bibinfo
  {author} {\bibfnamefont {A.}~\bibnamefont {Guttridge}},\ and\ \bibinfo
  {author} {\bibfnamefont {S.~L.}\ \bibnamefont {Cornish}},\ }\href@noop {}
  {\bibfield  {journal} {\bibinfo  {journal} {New Journal of Physics}\ }\textbf
  {\bibinfo {volume} {24}},\ \bibinfo {pages} {103022} (\bibinfo {year}
  {2022})}\BibitemShut {NoStop}%
\bibitem [{\citenamefont {Wang}\ \emph {et~al.}(2019)\citenamefont {Wang},
  \citenamefont {He}, \citenamefont {Guo}, \citenamefont {Xu}, \citenamefont
  {Sheng}, \citenamefont {Zhuang}, \citenamefont {Xiong}, \citenamefont {Liu},
  \citenamefont {Wang},\ and\ \citenamefont {Zhan}}]{wang2019preparation}%
  \BibitemOpen
  \bibfield  {author} {\bibinfo {author} {\bibfnamefont {K.}~\bibnamefont
  {Wang}}, \bibinfo {author} {\bibfnamefont {X.}~\bibnamefont {He}}, \bibinfo
  {author} {\bibfnamefont {R.}~\bibnamefont {Guo}}, \bibinfo {author}
  {\bibfnamefont {P.}~\bibnamefont {Xu}}, \bibinfo {author} {\bibfnamefont
  {C.}~\bibnamefont {Sheng}}, \bibinfo {author} {\bibfnamefont
  {J.}~\bibnamefont {Zhuang}}, \bibinfo {author} {\bibfnamefont
  {Z.}~\bibnamefont {Xiong}}, \bibinfo {author} {\bibfnamefont
  {M.}~\bibnamefont {Liu}}, \bibinfo {author} {\bibfnamefont {J.}~\bibnamefont
  {Wang}},\ and\ \bibinfo {author} {\bibfnamefont {M.}~\bibnamefont {Zhan}},\
  }\href@noop {} {\bibfield  {journal} {\bibinfo  {journal} {Physical Review
  A}\ }\textbf {\bibinfo {volume} {100}},\ \bibinfo {pages} {063429} (\bibinfo
  {year} {2019})}\BibitemShut {NoStop}%
\bibitem [{\citenamefont {Zhang}\ \emph {et~al.}(2020)\citenamefont {Zhang},
  \citenamefont {Yu}, \citenamefont {Cairncross}, \citenamefont {Wang},
  \citenamefont {Picard}, \citenamefont {Hood}, \citenamefont {Lin},
  \citenamefont {Hutson},\ and\ \citenamefont {Ni}}]{zhang2020forming}%
  \BibitemOpen
  \bibfield  {author} {\bibinfo {author} {\bibfnamefont {J.~T.}\ \bibnamefont
  {Zhang}}, \bibinfo {author} {\bibfnamefont {Y.}~\bibnamefont {Yu}}, \bibinfo
  {author} {\bibfnamefont {W.~B.}\ \bibnamefont {Cairncross}}, \bibinfo
  {author} {\bibfnamefont {K.}~\bibnamefont {Wang}}, \bibinfo {author}
  {\bibfnamefont {L.~R.}\ \bibnamefont {Picard}}, \bibinfo {author}
  {\bibfnamefont {J.~D.}\ \bibnamefont {Hood}}, \bibinfo {author}
  {\bibfnamefont {Y.-W.}\ \bibnamefont {Lin}}, \bibinfo {author} {\bibfnamefont
  {J.~M.}\ \bibnamefont {Hutson}},\ and\ \bibinfo {author} {\bibfnamefont
  {K.-K.}\ \bibnamefont {Ni}},\ }\href@noop {} {\bibfield  {journal} {\bibinfo
  {journal} {Physical Review Letters}\ }\textbf {\bibinfo {volume} {124}},\
  \bibinfo {pages} {253401} (\bibinfo {year} {2020})}\BibitemShut {NoStop}%
\bibitem [{\citenamefont {Holland}\ \emph {et~al.}(2022)\citenamefont
  {Holland}, \citenamefont {Lu},\ and\ \citenamefont
  {Cheuk}}]{holland2022demand}%
  \BibitemOpen
  \bibfield  {author} {\bibinfo {author} {\bibfnamefont {C.~M.}\ \bibnamefont
  {Holland}}, \bibinfo {author} {\bibfnamefont {Y.}~\bibnamefont {Lu}},\ and\
  \bibinfo {author} {\bibfnamefont {L.~W.}\ \bibnamefont {Cheuk}},\ }\href@noop
  {} {\bibfield  {journal} {\bibinfo  {journal} {arXiv preprint
  arXiv:2210.06309}\ } (\bibinfo {year} {2022})}\BibitemShut {NoStop}%
\bibitem [{\citenamefont {Bao}\ \emph {et~al.}(2022)\citenamefont {Bao},
  \citenamefont {Yu}, \citenamefont {Anderegg}, \citenamefont {Chae},
  \citenamefont {Ketterle}, \citenamefont {Ni},\ and\ \citenamefont
  {Doyle}}]{bao2022dipolar}%
  \BibitemOpen
  \bibfield  {author} {\bibinfo {author} {\bibfnamefont {Y.}~\bibnamefont
  {Bao}}, \bibinfo {author} {\bibfnamefont {S.~S.}\ \bibnamefont {Yu}},
  \bibinfo {author} {\bibfnamefont {L.}~\bibnamefont {Anderegg}}, \bibinfo
  {author} {\bibfnamefont {E.}~\bibnamefont {Chae}}, \bibinfo {author}
  {\bibfnamefont {W.}~\bibnamefont {Ketterle}}, \bibinfo {author}
  {\bibfnamefont {K.-K.}\ \bibnamefont {Ni}},\ and\ \bibinfo {author}
  {\bibfnamefont {J.~M.}\ \bibnamefont {Doyle}},\ }\href@noop {} {\bibfield
  {journal} {\bibinfo  {journal} {arXiv preprint arXiv:2211.09780}\ } (\bibinfo
  {year} {2022})}\BibitemShut {NoStop}%
\bibitem [{\citenamefont {Endres}\ \emph {et~al.}(2016)\citenamefont {Endres},
  \citenamefont {Bernien}, \citenamefont {Keesling}, \citenamefont {Levine},
  \citenamefont {Anschuetz}, \citenamefont {Krajenbrink}, \citenamefont
  {Senko}, \citenamefont {Vuletic}, \citenamefont {Greiner},\ and\
  \citenamefont {Lukin}}]{endres2016atom}%
  \BibitemOpen
  \bibfield  {author} {\bibinfo {author} {\bibfnamefont {M.}~\bibnamefont
  {Endres}}, \bibinfo {author} {\bibfnamefont {H.}~\bibnamefont {Bernien}},
  \bibinfo {author} {\bibfnamefont {A.}~\bibnamefont {Keesling}}, \bibinfo
  {author} {\bibfnamefont {H.}~\bibnamefont {Levine}}, \bibinfo {author}
  {\bibfnamefont {E.~R.}\ \bibnamefont {Anschuetz}}, \bibinfo {author}
  {\bibfnamefont {A.}~\bibnamefont {Krajenbrink}}, \bibinfo {author}
  {\bibfnamefont {C.}~\bibnamefont {Senko}}, \bibinfo {author} {\bibfnamefont
  {V.}~\bibnamefont {Vuletic}}, \bibinfo {author} {\bibfnamefont
  {M.}~\bibnamefont {Greiner}},\ and\ \bibinfo {author} {\bibfnamefont {M.~D.}\
  \bibnamefont {Lukin}},\ }\href@noop {} {\bibfield  {journal} {\bibinfo
  {journal} {Science}\ }\textbf {\bibinfo {volume} {354}},\ \bibinfo {pages}
  {1024} (\bibinfo {year} {2016})}\BibitemShut {NoStop}%
\bibitem [{\citenamefont {Barredo}\ \emph {et~al.}(2018)\citenamefont
  {Barredo}, \citenamefont {Lienhard}, \citenamefont {De~Leseleuc},
  \citenamefont {Lahaye},\ and\ \citenamefont
  {Browaeys}}]{barredo2018synthetic}%
  \BibitemOpen
  \bibfield  {author} {\bibinfo {author} {\bibfnamefont {D.}~\bibnamefont
  {Barredo}}, \bibinfo {author} {\bibfnamefont {V.}~\bibnamefont {Lienhard}},
  \bibinfo {author} {\bibfnamefont {S.}~\bibnamefont {De~Leseleuc}}, \bibinfo
  {author} {\bibfnamefont {T.}~\bibnamefont {Lahaye}},\ and\ \bibinfo {author}
  {\bibfnamefont {A.}~\bibnamefont {Browaeys}},\ }\href@noop {} {\bibfield
  {journal} {\bibinfo  {journal} {Nature}\ }\textbf {\bibinfo {volume} {561}},\
  \bibinfo {pages} {79} (\bibinfo {year} {2018})}\BibitemShut {NoStop}%
\bibitem [{\citenamefont {Liu}\ \emph {et~al.}(2018)\citenamefont {Liu},
  \citenamefont {Hood}, \citenamefont {Yu}, \citenamefont {Zhang},
  \citenamefont {Hutzler}, \citenamefont {Rosenband},\ and\ \citenamefont
  {Ni}}]{liu2018building}%
  \BibitemOpen
  \bibfield  {author} {\bibinfo {author} {\bibfnamefont {L.}~\bibnamefont
  {Liu}}, \bibinfo {author} {\bibfnamefont {J.}~\bibnamefont {Hood}}, \bibinfo
  {author} {\bibfnamefont {Y.}~\bibnamefont {Yu}}, \bibinfo {author}
  {\bibfnamefont {J.}~\bibnamefont {Zhang}}, \bibinfo {author} {\bibfnamefont
  {N.}~\bibnamefont {Hutzler}}, \bibinfo {author} {\bibfnamefont
  {T.}~\bibnamefont {Rosenband}},\ and\ \bibinfo {author} {\bibfnamefont
  {K.-K.}\ \bibnamefont {Ni}},\ }\href@noop {} {\bibfield  {journal} {\bibinfo
  {journal} {Science}\ }\textbf {\bibinfo {volume} {360}},\ \bibinfo {pages}
  {900} (\bibinfo {year} {2018})}\BibitemShut {NoStop}%
\bibitem [{\citenamefont {Yu}\ \emph {et~al.}(2018)\citenamefont {Yu},
  \citenamefont {Hutzler}, \citenamefont {Zhang}, \citenamefont {Liu},
  \citenamefont {Hood}, \citenamefont {Rosenband},\ and\ \citenamefont
  {Ni}}]{yu2018motional}%
  \BibitemOpen
  \bibfield  {author} {\bibinfo {author} {\bibfnamefont {Y.}~\bibnamefont
  {Yu}}, \bibinfo {author} {\bibfnamefont {N.~R.}\ \bibnamefont {Hutzler}},
  \bibinfo {author} {\bibfnamefont {J.~T.}\ \bibnamefont {Zhang}}, \bibinfo
  {author} {\bibfnamefont {L.~R.}\ \bibnamefont {Liu}}, \bibinfo {author}
  {\bibfnamefont {J.~D.}\ \bibnamefont {Hood}}, \bibinfo {author}
  {\bibfnamefont {T.}~\bibnamefont {Rosenband}},\ and\ \bibinfo {author}
  {\bibfnamefont {K.-K.}\ \bibnamefont {Ni}},\ }\href@noop {} {\bibfield
  {journal} {\bibinfo  {journal} {Physical Review A}\ }\textbf {\bibinfo
  {volume} {97}},\ \bibinfo {pages} {063423} (\bibinfo {year}
  {2018})}\BibitemShut {NoStop}%
\bibitem [{\citenamefont {Huang}\ \emph {et~al.}(2022)\citenamefont {Huang},
  \citenamefont {Covey}, \citenamefont {Gadway} \emph
  {et~al.}}]{huang2022gray}%
  \BibitemOpen
  \bibfield  {author} {\bibinfo {author} {\bibfnamefont {C.}~\bibnamefont
  {Huang}}, \bibinfo {author} {\bibfnamefont {J.~P.}\ \bibnamefont {Covey}},
  \bibinfo {author} {\bibfnamefont {B.}~\bibnamefont {Gadway}}, \emph
  {et~al.},\ }\href@noop {} {\bibfield  {journal} {\bibinfo  {journal}
  {Physical Review Research}\ }\textbf {\bibinfo {volume} {4}},\ \bibinfo
  {pages} {013240} (\bibinfo {year} {2022})}\BibitemShut {NoStop}%
\bibitem [{\citenamefont {Covey}\ \emph {et~al.}(2019)\citenamefont {Covey},
  \citenamefont {Madjarov}, \citenamefont {Cooper},\ and\ \citenamefont
  {Endres}}]{covey20192000}%
  \BibitemOpen
  \bibfield  {author} {\bibinfo {author} {\bibfnamefont {J.~P.}\ \bibnamefont
  {Covey}}, \bibinfo {author} {\bibfnamefont {I.~S.}\ \bibnamefont {Madjarov}},
  \bibinfo {author} {\bibfnamefont {A.}~\bibnamefont {Cooper}},\ and\ \bibinfo
  {author} {\bibfnamefont {M.}~\bibnamefont {Endres}},\ }\href@noop {}
  {\bibfield  {journal} {\bibinfo  {journal} {Physical review letters}\
  }\textbf {\bibinfo {volume} {122}},\ \bibinfo {pages} {173201} (\bibinfo
  {year} {2019})}\BibitemShut {NoStop}%
\bibitem [{\citenamefont {Parsons}\ \emph {et~al.}(2016)\citenamefont
  {Parsons}, \citenamefont {Mazurenko}, \citenamefont {Chiu}, \citenamefont
  {Ji}, \citenamefont {Greif},\ and\ \citenamefont
  {Greiner}}]{parsons2016site}%
  \BibitemOpen
  \bibfield  {author} {\bibinfo {author} {\bibfnamefont {M.~F.}\ \bibnamefont
  {Parsons}}, \bibinfo {author} {\bibfnamefont {A.}~\bibnamefont {Mazurenko}},
  \bibinfo {author} {\bibfnamefont {C.~S.}\ \bibnamefont {Chiu}}, \bibinfo
  {author} {\bibfnamefont {G.}~\bibnamefont {Ji}}, \bibinfo {author}
  {\bibfnamefont {D.}~\bibnamefont {Greif}},\ and\ \bibinfo {author}
  {\bibfnamefont {M.}~\bibnamefont {Greiner}},\ }\href@noop {} {\bibfield
  {journal} {\bibinfo  {journal} {Science}\ }\textbf {\bibinfo {volume}
  {353}},\ \bibinfo {pages} {1253} (\bibinfo {year} {2016})}\BibitemShut
  {NoStop}%
\bibitem [{\citenamefont {Brown}\ \emph {et~al.}(2017)\citenamefont {Brown},
  \citenamefont {Mitra}, \citenamefont {Guardado-Sanchez}, \citenamefont
  {Schau{\ss}}, \citenamefont {Kondov}, \citenamefont {Khatami}, \citenamefont
  {Paiva}, \citenamefont {Trivedi}, \citenamefont {Huse},\ and\ \citenamefont
  {Bakr}}]{brown2017spin}%
  \BibitemOpen
  \bibfield  {author} {\bibinfo {author} {\bibfnamefont {P.~T.}\ \bibnamefont
  {Brown}}, \bibinfo {author} {\bibfnamefont {D.}~\bibnamefont {Mitra}},
  \bibinfo {author} {\bibfnamefont {E.}~\bibnamefont {Guardado-Sanchez}},
  \bibinfo {author} {\bibfnamefont {P.}~\bibnamefont {Schau{\ss}}}, \bibinfo
  {author} {\bibfnamefont {S.~S.}\ \bibnamefont {Kondov}}, \bibinfo {author}
  {\bibfnamefont {E.}~\bibnamefont {Khatami}}, \bibinfo {author} {\bibfnamefont
  {T.}~\bibnamefont {Paiva}}, \bibinfo {author} {\bibfnamefont
  {N.}~\bibnamefont {Trivedi}}, \bibinfo {author} {\bibfnamefont {D.~A.}\
  \bibnamefont {Huse}},\ and\ \bibinfo {author} {\bibfnamefont {W.~S.}\
  \bibnamefont {Bakr}},\ }\href@noop {} {\bibfield  {journal} {\bibinfo
  {journal} {Science}\ }\textbf {\bibinfo {volume} {357}},\ \bibinfo {pages}
  {1385} (\bibinfo {year} {2017})}\BibitemShut {NoStop}%
\bibitem [{\citenamefont {Bergschneider}\ \emph {et~al.}(2018)\citenamefont
  {Bergschneider}, \citenamefont {Klinkhamer}, \citenamefont {Becher},
  \citenamefont {Klemt}, \citenamefont {Z{\"u}rn}, \citenamefont {Preiss},\
  and\ \citenamefont {Jochim}}]{bergschneider2018spin}%
  \BibitemOpen
  \bibfield  {author} {\bibinfo {author} {\bibfnamefont {A.}~\bibnamefont
  {Bergschneider}}, \bibinfo {author} {\bibfnamefont {V.~M.}\ \bibnamefont
  {Klinkhamer}}, \bibinfo {author} {\bibfnamefont {J.~H.}\ \bibnamefont
  {Becher}}, \bibinfo {author} {\bibfnamefont {R.}~\bibnamefont {Klemt}},
  \bibinfo {author} {\bibfnamefont {G.}~\bibnamefont {Z{\"u}rn}}, \bibinfo
  {author} {\bibfnamefont {P.~M.}\ \bibnamefont {Preiss}},\ and\ \bibinfo
  {author} {\bibfnamefont {S.}~\bibnamefont {Jochim}},\ }\href@noop {}
  {\bibfield  {journal} {\bibinfo  {journal} {Physical Review A}\ }\textbf
  {\bibinfo {volume} {97}},\ \bibinfo {pages} {063613} (\bibinfo {year}
  {2018})}\BibitemShut {NoStop}%
\bibitem [{\citenamefont {Holten}\ \emph {et~al.}(2021)\citenamefont {Holten},
  \citenamefont {Bayha}, \citenamefont {Subramanian}, \citenamefont {Heintze},
  \citenamefont {Preiss},\ and\ \citenamefont
  {Jochim}}]{holten2021observation}%
  \BibitemOpen
  \bibfield  {author} {\bibinfo {author} {\bibfnamefont {M.}~\bibnamefont
  {Holten}}, \bibinfo {author} {\bibfnamefont {L.}~\bibnamefont {Bayha}},
  \bibinfo {author} {\bibfnamefont {K.}~\bibnamefont {Subramanian}}, \bibinfo
  {author} {\bibfnamefont {C.}~\bibnamefont {Heintze}}, \bibinfo {author}
  {\bibfnamefont {P.~M.}\ \bibnamefont {Preiss}},\ and\ \bibinfo {author}
  {\bibfnamefont {S.}~\bibnamefont {Jochim}},\ }\href@noop {} {\bibfield
  {journal} {\bibinfo  {journal} {Physical Review Letters}\ }\textbf {\bibinfo
  {volume} {126}},\ \bibinfo {pages} {020401} (\bibinfo {year}
  {2021})}\BibitemShut {NoStop}%
\bibitem [{\citenamefont {Brown}\ \emph {et~al.}(2019)\citenamefont {Brown},
  \citenamefont {Thiele}, \citenamefont {Kiehl}, \citenamefont {Hsu},\ and\
  \citenamefont {Regal}}]{brown2019gray}%
  \BibitemOpen
  \bibfield  {author} {\bibinfo {author} {\bibfnamefont {M.}~\bibnamefont
  {Brown}}, \bibinfo {author} {\bibfnamefont {T.}~\bibnamefont {Thiele}},
  \bibinfo {author} {\bibfnamefont {C.}~\bibnamefont {Kiehl}}, \bibinfo
  {author} {\bibfnamefont {T.-W.}\ \bibnamefont {Hsu}},\ and\ \bibinfo {author}
  {\bibfnamefont {C.}~\bibnamefont {Regal}},\ }\href@noop {} {\bibfield
  {journal} {\bibinfo  {journal} {Physical Review X}\ }\textbf {\bibinfo
  {volume} {9}},\ \bibinfo {pages} {011057} (\bibinfo {year}
  {2019})}\BibitemShut {NoStop}%
\bibitem [{\citenamefont {Gr{\"u}nzweig}\ \emph {et~al.}(2010)\citenamefont
  {Gr{\"u}nzweig}, \citenamefont {Hilliard}, \citenamefont {McGovern},\ and\
  \citenamefont {Andersen}}]{grunzweig2010near}%
  \BibitemOpen
  \bibfield  {author} {\bibinfo {author} {\bibfnamefont {T.}~\bibnamefont
  {Gr{\"u}nzweig}}, \bibinfo {author} {\bibfnamefont {A.}~\bibnamefont
  {Hilliard}}, \bibinfo {author} {\bibfnamefont {M.}~\bibnamefont {McGovern}},\
  and\ \bibinfo {author} {\bibfnamefont {M.}~\bibnamefont {Andersen}},\
  }\href@noop {} {\bibfield  {journal} {\bibinfo  {journal} {Nature Physics}\
  }\textbf {\bibinfo {volume} {6}},\ \bibinfo {pages} {951} (\bibinfo {year}
  {2010})}\BibitemShut {NoStop}%
\bibitem [{\citenamefont {Lester}\ \emph {et~al.}(2015)\citenamefont {Lester},
  \citenamefont {Luick}, \citenamefont {Kaufman}, \citenamefont {Reynolds},\
  and\ \citenamefont {Regal}}]{lester2015rapid}%
  \BibitemOpen
  \bibfield  {author} {\bibinfo {author} {\bibfnamefont {B.~J.}\ \bibnamefont
  {Lester}}, \bibinfo {author} {\bibfnamefont {N.}~\bibnamefont {Luick}},
  \bibinfo {author} {\bibfnamefont {A.~M.}\ \bibnamefont {Kaufman}}, \bibinfo
  {author} {\bibfnamefont {C.~M.}\ \bibnamefont {Reynolds}},\ and\ \bibinfo
  {author} {\bibfnamefont {C.~A.}\ \bibnamefont {Regal}},\ }\href@noop {}
  {\bibfield  {journal} {\bibinfo  {journal} {Physical review letters}\
  }\textbf {\bibinfo {volume} {115}},\ \bibinfo {pages} {073003} (\bibinfo
  {year} {2015})}\BibitemShut {NoStop}%
\bibitem [{\citenamefont {Cheuk}\ \emph {et~al.}(2018)\citenamefont {Cheuk},
  \citenamefont {Anderegg}, \citenamefont {Augenbraun}, \citenamefont {Bao},
  \citenamefont {Burchesky}, \citenamefont {Ketterle},\ and\ \citenamefont
  {Doyle}}]{cheuk2018lambda}%
  \BibitemOpen
  \bibfield  {author} {\bibinfo {author} {\bibfnamefont {L.~W.}\ \bibnamefont
  {Cheuk}}, \bibinfo {author} {\bibfnamefont {L.}~\bibnamefont {Anderegg}},
  \bibinfo {author} {\bibfnamefont {B.~L.}\ \bibnamefont {Augenbraun}},
  \bibinfo {author} {\bibfnamefont {Y.}~\bibnamefont {Bao}}, \bibinfo {author}
  {\bibfnamefont {S.}~\bibnamefont {Burchesky}}, \bibinfo {author}
  {\bibfnamefont {W.}~\bibnamefont {Ketterle}},\ and\ \bibinfo {author}
  {\bibfnamefont {J.~M.}\ \bibnamefont {Doyle}},\ }\href@noop {} {\bibfield
  {journal} {\bibinfo  {journal} {Physical review letters}\ }\textbf {\bibinfo
  {volume} {121}},\ \bibinfo {pages} {083201} (\bibinfo {year}
  {2018})}\BibitemShut {NoStop}%
\bibitem [{\citenamefont {Tuchendler}\ \emph {et~al.}(2008)\citenamefont
  {Tuchendler}, \citenamefont {Lance}, \citenamefont {Browaeys}, \citenamefont
  {Sortais},\ and\ \citenamefont {Grangier}}]{tuchendler2008energy}%
  \BibitemOpen
  \bibfield  {author} {\bibinfo {author} {\bibfnamefont {C.}~\bibnamefont
  {Tuchendler}}, \bibinfo {author} {\bibfnamefont {A.~M.}\ \bibnamefont
  {Lance}}, \bibinfo {author} {\bibfnamefont {A.}~\bibnamefont {Browaeys}},
  \bibinfo {author} {\bibfnamefont {Y.~R.}\ \bibnamefont {Sortais}},\ and\
  \bibinfo {author} {\bibfnamefont {P.}~\bibnamefont {Grangier}},\ }\href@noop
  {} {\bibfield  {journal} {\bibinfo  {journal} {Physical Review A}\ }\textbf
  {\bibinfo {volume} {78}},\ \bibinfo {pages} {033425} (\bibinfo {year}
  {2008})}\BibitemShut {NoStop}%
\bibitem [{sup()}]{supp}%
  \BibitemOpen
  \href@noop {} {}\bibinfo {howpublished}
  {\url{URL_will_be_inserted_by_publisher}}\BibitemShut {NoStop}%
\bibitem [{\citenamefont {Salomon}\ \emph {et~al.}(2014)\citenamefont
  {Salomon}, \citenamefont {Fouch{\'e}}, \citenamefont {Wang}, \citenamefont
  {Aspect}, \citenamefont {Bouyer},\ and\ \citenamefont
  {Bourdel}}]{salomon2014gray}%
  \BibitemOpen
  \bibfield  {author} {\bibinfo {author} {\bibfnamefont {G.}~\bibnamefont
  {Salomon}}, \bibinfo {author} {\bibfnamefont {L.}~\bibnamefont {Fouch{\'e}}},
  \bibinfo {author} {\bibfnamefont {P.}~\bibnamefont {Wang}}, \bibinfo {author}
  {\bibfnamefont {A.}~\bibnamefont {Aspect}}, \bibinfo {author} {\bibfnamefont
  {P.}~\bibnamefont {Bouyer}},\ and\ \bibinfo {author} {\bibfnamefont
  {T.}~\bibnamefont {Bourdel}},\ }\href@noop {} {\bibfield  {journal} {\bibinfo
   {journal} {Europhysics Letters}\ }\textbf {\bibinfo {volume} {104}},\
  \bibinfo {pages} {63002} (\bibinfo {year} {2014})}\BibitemShut {NoStop}%
\end{thebibliography}%


%apsrev4-2.bst 2019-01-14 (MD) hand-edited version of apsrev4-1.bst
%Control: key (0)
%Control: author (72) initials jnrlst
%Control: editor formatted (1) identically to author
%Control: production of article title (-1) disabled
%Control: page (0) single
%Control: year (1) truncated
%Control: production of eprint (0) enabled
\providecommand{\noopsort}[1]{}\providecommand{\singleletter}[1]{#1}%
\begin{thebibliography}{2}%
\makeatletter
\providecommand \@ifxundefined [1]{%
 \@ifx{#1\undefined}
}%
\providecommand \@ifnum [1]{%
 \ifnum #1\expandafter \@firstoftwo
 \else \expandafter \@secondoftwo
 \fi
}%
\providecommand \@ifx [1]{%
 \ifx #1\expandafter \@firstoftwo
 \else \expandafter \@secondoftwo
 \fi
}%
\providecommand \natexlab [1]{#1}%
\providecommand \enquote  [1]{``#1''}%
\providecommand \bibnamefont  [1]{#1}%
\providecommand \bibfnamefont [1]{#1}%
\providecommand \citenamefont [1]{#1}%
\providecommand \href@noop [0]{\@secondoftwo}%
\providecommand \href [0]{\begingroup \@sanitize@url \@href}%
\providecommand \@href[1]{\@@startlink{#1}\@@href}%
\providecommand \@@href[1]{\endgroup#1\@@endlink}%
\providecommand \@sanitize@url [0]{\catcode `\\12\catcode `\$12\catcode
  `\&12\catcode `\#12\catcode `\^12\catcode `\_12\catcode `\%12\relax}%
\providecommand \@@startlink[1]{}%
\providecommand \@@endlink[0]{}%
\providecommand \url  [0]{\begingroup\@sanitize@url \@url }%
\providecommand \@url [1]{\endgroup\@href {#1}{\urlprefix }}%
\providecommand \urlprefix  [0]{URL }%
\providecommand \Eprint [0]{\href }%
\providecommand \doibase [0]{https://doi.org/}%
\providecommand \selectlanguage [0]{\@gobble}%
\providecommand \bibinfo  [0]{\@secondoftwo}%
\providecommand \bibfield  [0]{\@secondoftwo}%
\providecommand \translation [1]{[#1]}%
\providecommand \BibitemOpen [0]{}%
\providecommand \bibitemStop [0]{}%
\providecommand \bibitemNoStop [0]{.\EOS\space}%
\providecommand \EOS [0]{\spacefactor3000\relax}%
\providecommand \BibitemShut  [1]{\csname bibitem#1\endcsname}%
\let\auto@bib@innerbib\@empty
%</preamble>
\bibitem [{\citenamefont {Steck}(2023)}]{steck}%
  \BibitemOpen
  \bibfield  {author} {\bibinfo {author} {\bibfnamefont {D.~A.}\ \bibnamefont
  {Steck}},\ }\href@noop {} {\bibinfo {title} {Quantum and atom optics,
  available online at http://steck.us/teaching (revision 0.13.19, 6 april
  2023)}} (\bibinfo {year} {2023})\BibitemShut {NoStop}%
\bibitem [{\citenamefont {Gehm}()}]{gehm2003}%
  \BibitemOpen
  \bibfield  {author} {\bibinfo {author} {\bibfnamefont {M.~E.}\ \bibnamefont
  {Gehm}},\ }\href@noop {} {\bibinfo {title} {Properties of {$^6$Li}}},\
  \bibinfo {howpublished}
  {\url{https://jet.physics.ncsu.edu/techdocs/pdf/PropertiesOfLi.pdf}},\
  \bibinfo {note} {accessed: 2023-04-28}\BibitemShut {NoStop}%
\end{thebibliography}%

\end{document}

\title{Supplementary Material}
\maketitle
\section{\label{sec:Experimental Data}Experimental Data}
\begin{figure}[h]
    \centering
    \includegraphics[width=.35\textwidth]{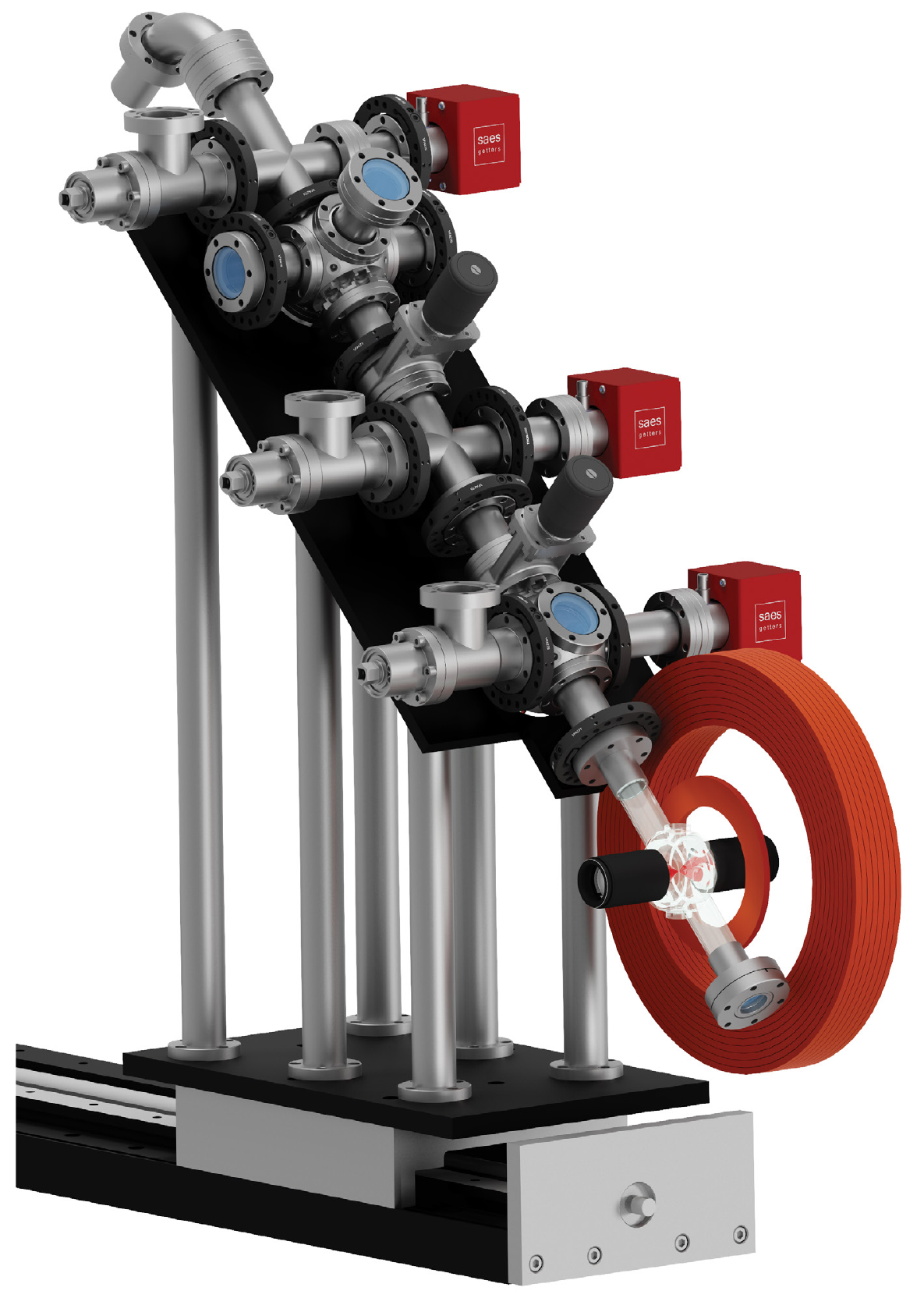}
    \caption{\label{fig:FullChamber}Full view of our vacuum chamber, which is oriented at 45$^{\circ}$ with respect to the optics table. Additionally, the entire chamber is mounted on a translational stage. These features allow for convenient optical access and beam alignment at the atoms. The lithium-6 effusive oven, located at the back of the chamber, is separated from the double-sided glass octagon cell by two stages of differential pumping (marked by the three red Saes ion/getter pumps). The optically accessible 4-way cross just after the effusive oven is useful for characterizing and shaping the atomic beam. Shim coils, and the front MOT and mini MOT coils are removed for clarity.  }
\end{figure}

Figure~\ref{fig:FullChamber} presents our experimental chamber in its entirety, with all of the shim coils and the front MOT and mini-MOT coils removed for clarity. The science region of our vacuum chamber is comprised of a double-sided octagon glass cell (Precision Glassblowing) treated with nano-texture etching (TelAztec). Our MOT beams are oriented such that the 2 radial beams are orthogonal in the radial plane, while the axial beam is 30$^{\circ}$ off axis. Due to Lithium's low vapor pressure, we introduce atoms into the gas-phase by use of an atomic beam source. A stainless steel oven containing 14g of isotopically enriched $^6$Li is heated to 300$^{\circ}$ Celsius with band heaters. A nozzle comprised of a triangular array of 21 high aspect ratio stainless steel microcapillary tubes couples the lithium oven to the ultra-high vacuum chamber, thereby forming an effusive beam of $^6$Li atoms that delivers a flux of $\sim$10$^9$ atoms/s into the science region of our chamber.

We utilize two stages of differential pumping by separating the three regions of our vacuum chamber with two small stainless steel tubes, both of which have internal diameters that are twice that necessitated by the aspect ratio of the atomic beam. Atoms are decelerated via a Zeeman slower laser that is passed through the far-end of the octagon glass cell via a sapphire viewport. Laser light that addresses the D1 and D2 lines of Lithium are generated by two external cavity diode lasers (Toptica DL pros). The D1 laser is locked to the D1 central crossover transition via saturated absorption spectroscopy and the D2 laser is offset phase-locked to the D1 laser. Repumper light for D2 MOT loading and D1 cooling is generated with a free-space electro-optic modulator (Qubig).

To circumvent considerable eddy currents after the main MOT coils are switched off, a pair of smaller mini-MOT coils remain on for 50 ms, after which time all residual magnetic fields are switched off and $\Lambda$-enhanced gray molasses ($\Lambda$-GM) is performed on the D1 line of the bulk gas, reducing the temperature of the atoms from 1 mK to 100 $\mu$K. Atoms are then probabilistically loaded into a 1064 nm optical tweezer, which is generated from an ultralow noise high power fiber laser (Azurlight Systems, 130 W). The 1064 nm light is coupled to the experiment via a polarization-maintaining optical fiber, and is split into an arbitrary array of individual trap sites using an acouso-optic deflector (IntraAction Corp) driven by an arbitrary waveform generator (16bit 625MS/s M4i6622-x8). The tweezer light is then passed through a dichroic filter and then through a high numerical aperture objective (Special Optics, 0.55 NA), through which it is focused down to approximately 1.0 $\mu$m at the center of the glass cell.

\begin{figure}
    \centering
    \includegraphics[width=.43\textwidth]{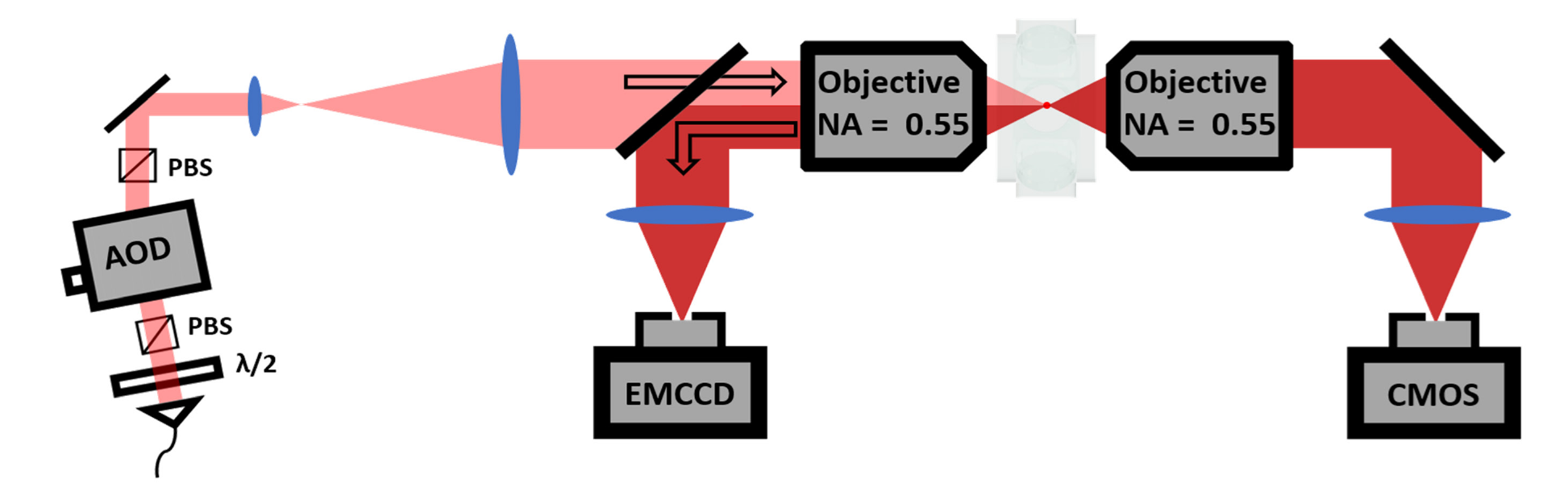}
    \caption{\label{fig:OpticsCartoon} Cartoon schematic of atom imaging. 1064 nm tweezer light is filtered and split into an array by an acousto-optic modulator. The light is expanded by a factor of ten and focused down through an objective to the center of the glass cell. Scattered photons from trapped atoms are collected back through the same objective, bounced off of a dichroic mirror and imaged onto an EMCCD camera. Additionally, a second back objective is used to image both the tweezer sites and the MOT onto a CMOS camera. }
\end{figure}
%\FloatBarrier

\begin{figure}
    \centering
    \includegraphics[width=.33\textwidth]{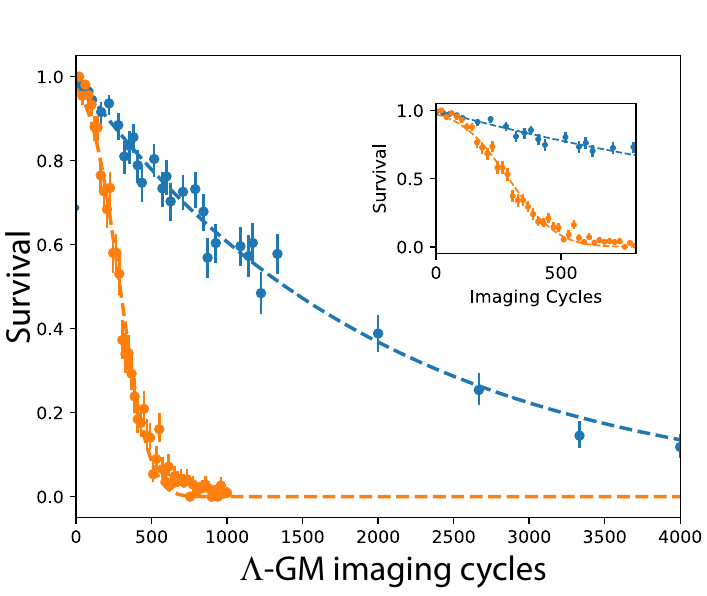}
    \caption{\label{fig:Survival_erf}Atom survival vs. number of images. Inset shows early timescale. Each imaging cycle lasts 15 ms. }
\end{figure}

%\FloatBarrier

%\begin{comment}
%We calculate the off resonant, inelastic scattering rate from the 165 THz %detuned, 3.4 mK deep tweezer to be $\sim$10 Hz. While not a high rate, it %appears likely that when combined with the large recoil energy of $^6$Li %(3.5$\mu$K), the thermal distribution of the atom in the tweezer rapidly %expands further and further outside of the trap, leading to atom loss. 300 %imaging cycles = 4.5 s...
%\end{comment}

The survival plots for continuous $\Lambda$-GM (blue) and simply holding in the tweezer (orange) are reproduced in Figure~\ref{fig:Survival_erf}. As mentioned in the main text, the survival decay curve of simply holding the atom in the tweezer took the form of a sigmoid, rather than an exponential, function.  We fit the tweezer hold time decay curve to an error function of a normalized Gaussian with zero offset. The inset of the plot shows early timescales, where the goodness of the fit is apparent. The good agreement implies that that probability density function (PDF) of the time required to heat the atom out of the trap is, to a good approximation, a Gaussian function. By a simple conversion with $\Gamma$, the PDF of the number of photons required to heat the atom out of the trap is also a Gaussian function. We can see from Figure~\ref{fig:Survival_erf}  that holding the atom in the tweezer for the time equivalent of taking 300 $\Lambda$-GM images knocks the atom out of the trap with 50$\%$ likelihood. This corresponds to a tweezer hold time of 4.5 s. Using an off-resonant scattering rate of 10 Hz implies that a lithium atom in the tweezer, initially at 100 $\mu$K, needs to only scatter 45 photons before 50 $\%$ likelihood of heating out of the trap. 

\FloatBarrier

\begin{figure}[h]
    \centering
    \includegraphics[width=0.35\textwidth]{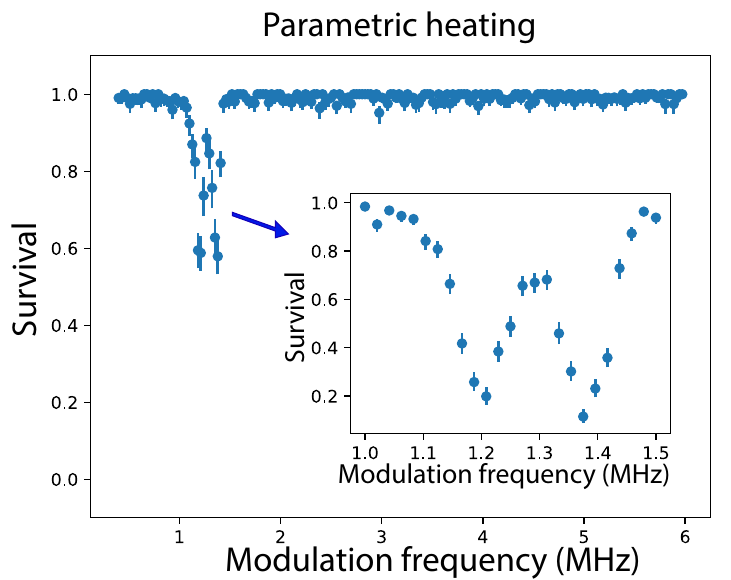}
    \caption{\label{fig:ParametricHeating} Atom survival vs. trap depth modulation frequency. Lithium's light mass leads to relatively high radial trap frequencies of 600 and 700 kHz. }
\end{figure}

\begin{figure}[h]
    \centering
    \includegraphics[width=0.66\textwidth]{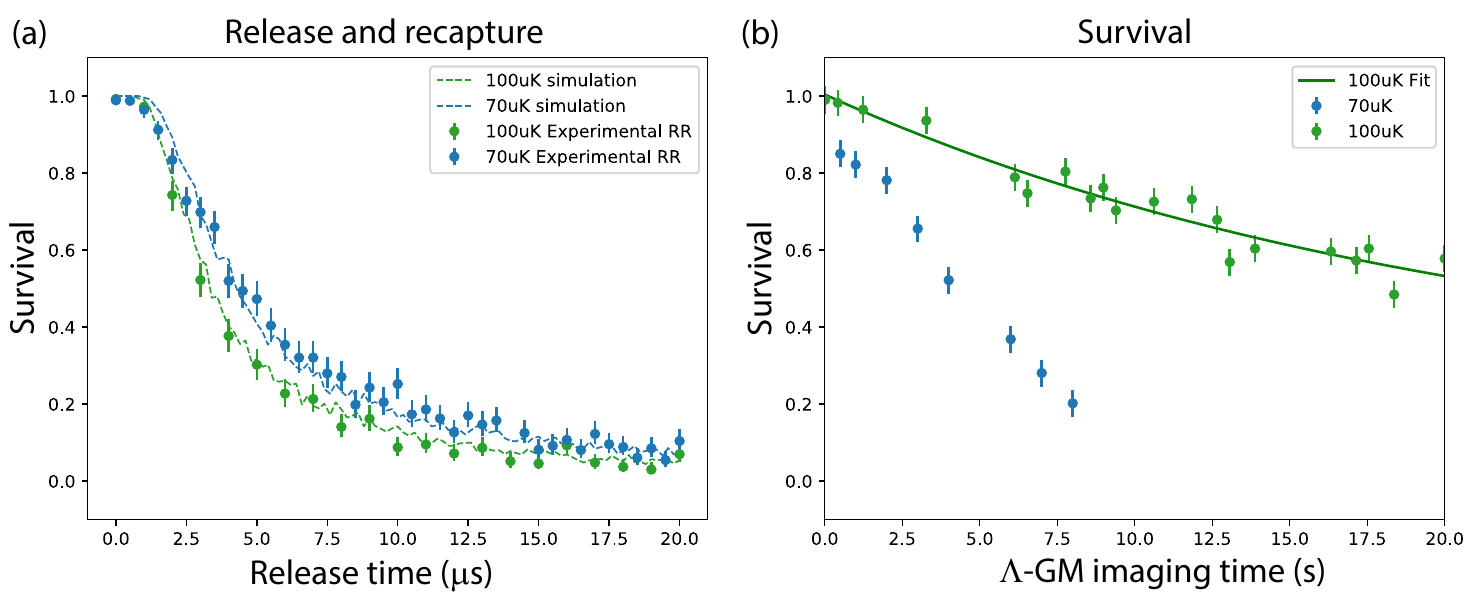}
    \caption{\label{fig:100_70_RR_Survival} These plots highlight the sensitivity of $\Lambda$-GM parameters on repeated imaging survival. At $\Lambda$-GM conditions corresponding to the lowest achieved temperature (measured via release and recapture (a)) the atom survival under constant application of the cooling light is much worse than that of the $\Lambda$-GM optimized for 100 $\mu$K (b). As mentioned in the main text, the survival $1/e$ lifetime of the 70 $\mu$K data is nearly identical to that of simply holding the atom in the tweezer.}
\end{figure}

\begin{figure}[h]
    \centering
    \includegraphics[width=0.66\textwidth]{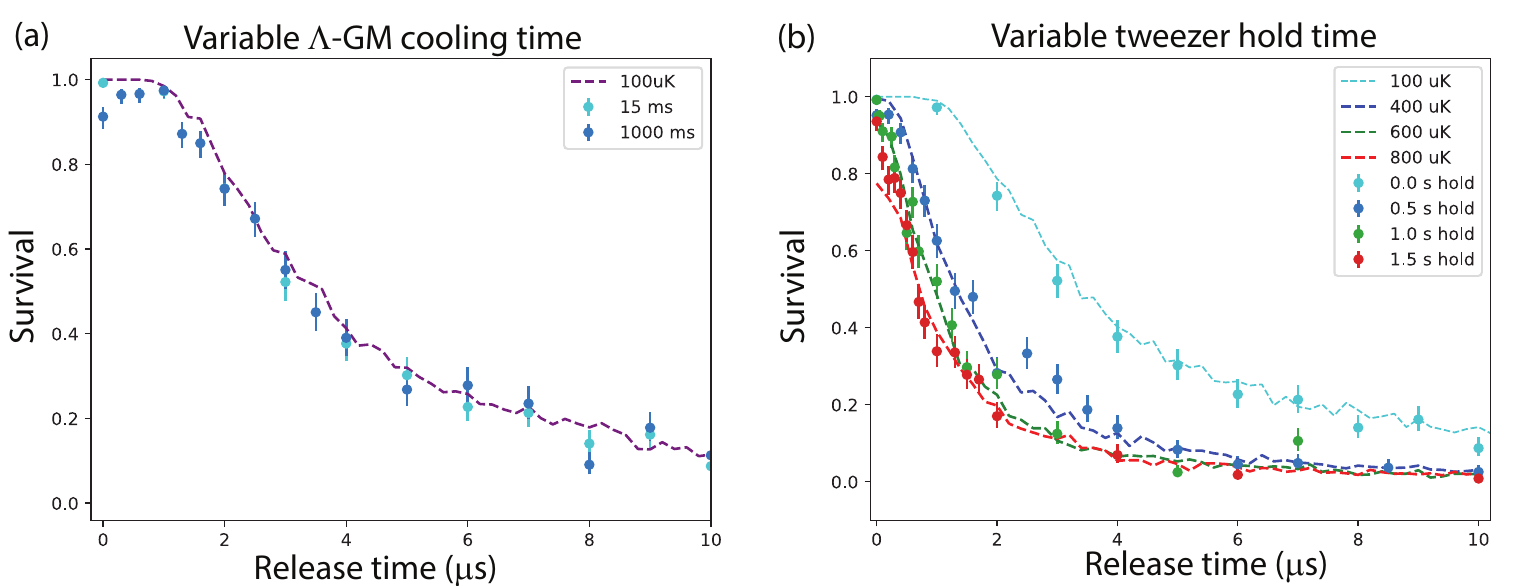}
    \caption{\label{fig:GM_vs_Hold_Survival}The above two plots show the time-dependant effects of application of cooling light vs. application of no light in the tweezer. In the first case (a) we see that the temperature of the atom is the same after 15 ms and 1,000 ms of applied $\Lambda$-GM light. In the second case (b), we see that the atom heats rapidly by off-resonant scattering of the trap light. }
\end{figure}

%\FloatBarrier

%%%%%%%%%%%%%%%%%%%%%%%%%%  THEORY %%%%%%%%%%%%%%%%%%%%%%%%%%%%%%%%%%%
\section{\label{Cooling simulation}Tweezer cooling simulation}

\begin{figure}[h!]
    \centering
    \includegraphics[width=0.3\textwidth]{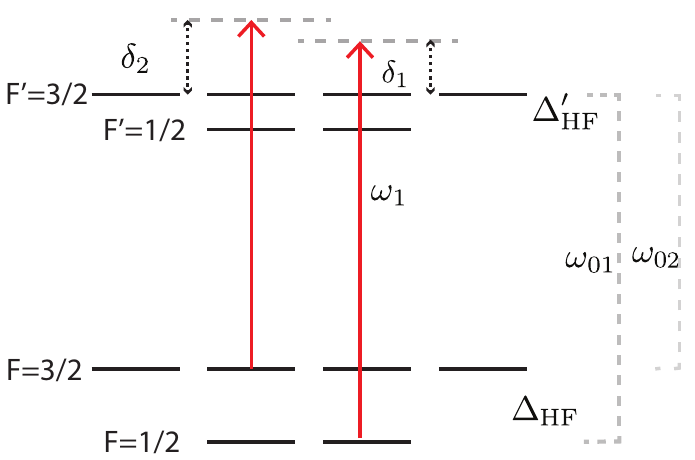}
    \caption{Lithium 6 D1 level structure and definition of frequencies, detunings, and hyperfine splittings.  }
    \label{fig:Li6_levels}
\end{figure}

The tweezer cooling simulation is performed by solving the steady state master equation for the D1 levels of lithium 6 in a harmonic oscillator.  We simulate up to 14 harmonic levels in 1D and obtain accurate predictions for temperature and scattering rate.   We also develop a new picture for cooling in gray molasses (GM) and $\Lambda$-enhanced GM in and near the Lamb-Dicke regime, which leads to a  simple criteria for efficient cooling: the bright state AC stark shift must be similar to the trapping frequency energy $\hbar \omega$. In a future paper, we will show how this picture can also describe polarization gradient (PG), Doppler cooling, and sideband cooling.  

%The primary challenge in theoretical modeling is constructing the total Hamiltonian, for a multilevel atomic system in the correct basis, before actually simulating it in the code.
%Our system comprises a lithium-6 atom confined within a 1 µm-sized trap, generated by a 1064 nm Gaussian-profile laser with a trap depth of several mK. The trap exhibits a nearly harmonic potential, and this trapped atom is also exposed to two 4 mm coherent lasers, incident along all three axes in a $\sigma_+ \sigma_-$ configuration for cooling.
%The ground and excited states experience distinct tweezer potentials due to their different polarizabilities. 

%To derive the total Hamiltonian (H), we write the bare atom Hamiltonian ($H_\text{0}$), which encompasses the ground and excited states, along with their corresponding hyperfine levels. Next, we represent the tweezer motional states using the harmonic oscillator Hamiltonian ($H_\text{HO}$) and integrate the interaction Hamiltonian ($H_\text{AF}$) to account for atom-light coupling, energy shifts, and laser-induced transitions. The constructed total H is used for the Grey Molasses cooling simulation, which involves determining the master equation's steady state while considering the hyperfine structure of the Li atom, specifically F = 1/2 and F = 3/2 for the ground state, and F' = 1/2 and F' = 3/2 for the excited state during the D1 transition. Both cooling and repumper light, as illustrated in Fig S7, are taken into account.

\subsubsection{Master Equation Derivation}
Before diving into the derivation, first we work out the dipole moment operator and electric field operator in the preferred basis.  The atom-dipole notation follows the conventions of Ref.~\cite{steck}. 
The dipole operator is expressed in spherical vector form as
\begin{equation}
\mathbf{d} = \sum_q d_q \hat{\mathbf{e}}_q^*,
\end{equation}
with components $d_{\pm1} = \mp \frac{1}{\sqrt{2}} (d_x \pm i d_y)$ and $d_0 = d_z$. The unit spherical vectors are similarly defined as $\hat{\mathbf{e}}_{\pm1} = \mp \frac{1}{\sqrt{2}} (\hat{\mathbf{e}}_x \pm i \hat{\mathbf{e}}_y)$ and $\hat{\mathbf{e}}_0 = \hat{\mathbf{e}}_z$. Spherical vectors possess properties such as $(d_q)^* = (-1)^q d_{-q}$ and dot product $\mathbf{A} \cdot \mathbf{B} = \sum_q (-1)^q A_q B_{-q} = \sum_q A_q B_q^*$, with orthogonality $\hat{e}_q \cdot \hat{e}_{q'}^* = \delta_{qq'}$.

The identity operator in the atom basis is a sum of all ground and D1 excited states:
\begin{equation}
\hat{\mathcal{I}} = \sum_{F_g m_g} |F_g m_g \rangle \langle F_g m_g | + \sum_{F_e m_e} |F_e m_e \rangle \langle F_e m_e |.
\end{equation}
Multiplying the dipole operator on both sides by the identity $\hat{\mathcal{I}} d_q \hat{\mathcal{I}}$ and using the Wigner-Eckert theorem and angular momentum identities yields an expression for the dipole momentum operator in terms of a collective lowering operator $\Sigma_q$~\cite{steck}:
\begin{equation}
d_q = \langle J_g || d || J_e \rangle \left(\Sigma_q + (-1)^q \Sigma_{-q}^\dag \right) = d_q^{(+)} + d_q^{(-)}.
\end{equation}
We further separate the collective lowering operator into a lowering operator for both ground states:
\begin{equation}
\Sigma_q = \sum_{F_g} \Sigma_{q, F_g}, 
\end{equation}
with lowering operator to a single hyperfine ground state as:
\begin{equation}
\Sigma_{q, F_g} = \sum_{m_g F_e m_e} (-1)^{F_e + J_g + I + 1} \sqrt{(2F_e+1)(2 J_g + 1)} \langle F_g m_g|F_e m_e ; 1 q \rangle
 \begin{Bmatrix} J_e & J_g & 1 \\ F_g & F_e & 1 \end{Bmatrix}
|F_g m_g \rangle \langle F_e m_e|.
\end{equation}
This lowering operator lowers all excited states to the $F_g$ ground state while changing $m$ by $q$.

%\subsubsection{Electric field operator}
The electric field operator written in complex form with a rotating $(+)$ and counter rotating  $(-)$ terms is 
\begin{equation}
\mathbf{E}(\mathbf{r} , t) = \mathbf{E}^{(+)}(\mathbf{r} , t) + \mathbf{E}^{(-)}(\mathbf{r} , t) =  \sum_i  \left(  \boldsymbol{\mathcal{E}}^{(+)}_i(\mathbf{r})  e^{- i \omega_i t}  + \text{c.c.} \right)   
\end{equation}
Here $i$ denotes the frequency of the laser that addresses either the  $F=3/2$ or $F=1/2$ transition.  These frequencies are typically detuned relative to each by the ground state hyperfine splitting in order to make a $\Lambda$ system. 
The complex spatial field for a particular frequency $i$ is expressed in terms of a sum of plane waves
\begin{equation}
  \boldsymbol{\mathcal{E}}^{(+)}_i(\mathbf{r}) = \sum_j  \boldsymbol{\epsilon}_{ij} \frac{E_{ij}}{2} e^{i \mathbf{k}_{ij} \cdot \mathbf{r}}  
\end{equation}
The index $j$ denotes the beam index for one of the frequencies, $\mathbf{k}_{ij}$ is the wave-vector for frequency $i$ and beam $j$,  $\boldsymbol{\epsilon}_{ij}$ is the polarization vector,  and $E_{ij}$ is the complex electric field amplitude.  The intensity is related to the amplitude by $I = \frac{c \epsilon_0}{2} |E|^2 $.

%\subsubsection{Hamiltonian}

The Hamiltonian for the interaction between the atom and the electric field of the laser in the dipole approximation and rotating wave approximation (but not in the rotating frame) is
\begin{align}
    H_{\text{AF}} &= -\mathbf{d}^{(+)} \cdot \mathbf{E}^{(-)} - \mathbf{d}^{(-)} \cdot \mathbf{E}^{(+)} \nonumber \\
    &= -\sum_q (-1)^q \left( d_q^{(+)} E_{-q}^{(-)} + d_q^{(-)} E_{-q}^{(+)}  \right)  
\end{align}
We define the position-dependent but time-independent Rabi frequencies
\begin{equation}
    \Omega_{q,i}(\mathbf{r}) = -2 \langle J_g || \mathbf{d} || J_e \rangle  \frac{1}{\hbar}  \mathcal{E}^{(+)}_{q,i}(\mathbf{r})
\end{equation}
The atom-field interaction is then defined in terms of a sum over the two laser frequencies and all three polarizations 
\begin{equation}
    H_{AF} = \frac{1}{2} \sum_{q,i} \left( \Omega_{q,i}(\mathbf{r}) \Sigma_q^{(+)} e^{-i\omega_i t} + h.c. \right)
\end{equation}
% labels for the F=3/2 and F=1/2 ground states.  

We now go to the rotating frame of the lasers to get rid of the time-dependence.  Because the cooling light is mostly addressing the $F=\Fthree$ state and the repumper is mostly addressing the $F=1/2$ state, we put the two hyperfine ground states in the rotating frame of their respective laser by applying a unitary transformation
\begin{equation}
   U =  e^{- i \omega_{\Fthree} t   \sum_{m_g} |\Fthree,  m_g \rangle \langle \Fthree,  m_g |  }  e^{- i \omega_{\Fone} t   \sum_{m_g} |\Fone,  m_g \rangle \langle \Fone,  m_g  |  }   
\end{equation}
The new Hamiltonian transforms to $ H \rightarrow   U H U^\dag + i \hbar (\frac{\delta}{\delta t} U) U^\dag$ and the operators transform to  $ \sigma  \rightarrow U \sigma U^\dag$. 
From now on, all operators will be assumed to be in the rotating frame defined by this unitary operator.

In this rotating frame, the non-interacting Hamiltonian is (defining $F'=3/2$ excited state as zero energy) 
\begin{equation}
    \begin{split}  
      H_0 &=  \sum_{m_g} \big[ \delta_{1/2} |\Fone, m_g \rangle \langle \Fone, m_g|  + \delta_{3/2} |\Fthree, m_g \rangle \langle \Fthree, m_g|\big] \\
      &+ \sum_{m_e} \big[ - \Delta'_{\text{HF}} | \Fone, m_e\rangle \langle  \Fone, m_e |  + 0 \times |3/2, m_e \rangle \langle 3/2, me| \big]
    \end{split}
\end{equation}
As defined in the level diagram, $\Delta'_{HF}$ is the D1 excited state hyperfine splitting and is positive(negative) for GM(PG) cooling.

The atom-field interaction has two terms.  One is time-independent, where the cooling light only addresses the $F=3/2$ state and repumper light only addresses the $F=1/2$ state:
\begin{equation}
    H_{\text{AF}} = \frac{1}{2} \sum_q \left[  \Omega_{q, \Fthree} (\mathbf{r}) \Sigma^\dag_{q,\Fthree}  + \Omega_{q, \Fone} (\mathbf{r}) \Sigma^\dag_{q,\Fone}  + h.c.   \right] 
\end{equation}
The other atom field interaction is the cross-coupling terms consisting of the $F=3/2$ laser addressing the $F=1/2$ transition and $F=1/2$ laser addressing the $F=3/2$ transition.  These terms still have time-dependence in the rotating frame: 
\begin{equation}
 H_{\text{AF}}^\text{cross} = \sum_q \left[  \Omega_{q, \Fone} (\mathbf{r})  e^{i (\omega_{\Fthree} - \omega_{\Fone} )t} \Sigma^\dag_{q,\Fthree}  + \Omega_{q, \Fone} (\mathbf{r})  e^{i (\omega_{\Fone} - \omega_{\Fthree} )t} \Sigma^\dag_{q,\Fone}   + h.c. \right]   
\end{equation}
This cross-coupling will be treated later as a perturbation, which is a good approximation when $\delta \ll \Delta_{HF}$.  Since the ground state hyperfine splitting is several times larger than the detunings, they are off-resonant.

%\subsubsection{Harmonic states}

Lastly, the interaction with motional states is incorporated by first adding a harmonic oscillator Hamiltonian. 
\begin{equation}
H_\text{HO} =  \hbar \omega (a^\dag a  + 1/2 )
\end{equation}
where $\omega$ is the trapping frequency. Second, we promote the positions in $\Omega_{q,i}(\mathbf{r})$ to operators.
In our 1D simulation, we assume that the incoming light is propagating in the $z$ direction as $E_0 e^{ikz}$ with real amplitude $E_0$. We also assume there is a retro-reflected beam with the same magnitude $E_0 e^{i \phi }e^{ikz}$ but with a phase determined by the mirror position. We also assume that the polarization is in the $x$-$y$ plane. We define recoil operators 
\begin{equation}
    \hat{\mathcal{R}} = e^{i k \hat{z}}, \quad  \hat{\mathcal{R}}^\dag = e^{-i k \hat{z}}
\end{equation}
which are momentum displacement operators corresponding to a recoil from photon absorption or emission. 
We do not assume the Lamb-Dicke (LD) regime in our simulations, but we find it useful to expand the recoil operator in the LD regime to see what the dominant terms correspond to. The LD regime here would correspond to the expansion
\begin{equation}
    \hat{\mathcal{R}} \approx  1 + i k \hat{z} = 1 + i \eta_{LD} (\hat{a} + \hat{a}^\dag)
\end{equation}
where $\eta_{LD} = k z_0 $ and $z_0 = \sqrt{\hbar/2m\omega}$.  The first term corresponds to no motional state change, and the second term to either the motional state gaining or loosing one quanta.

In terms of the recoil operators, the complex electric field can be expressed as 
\begin{equation}
   \Omega_{q,i}( \hat{z}) = \Omega_i \left[ \epsilon_{q, \rightarrow} \hat{\mathcal{R}} + e^{i \phi_i} \epsilon_{q, \leftarrow} \hat{\mathcal{R}}^\dag \right]
\end{equation}
with scalar Rabi frequency 
\begin{equation}
    \Omega_i = -\frac{2}{\hbar} \langle J_g || \mathbf{d} || J_e \rangle \frac{E_{0,i}}{2}
 \end{equation}
%The arrow denotes whether the beam is going to the right (positive $z$) or left (negative $z$).
The arrows denote the direction of the beam. 

The atom-field interaction that couples the ground and excited state and motional states for a pair of counter-propagating beams is 
\begin{equation}
    H_{AF} = \sum_{q,i} \frac{\Omega_i}{2}  \left[ (    \epsilon_{q, \rightarrow} \hat{\mathcal{R}} + e^{i \phi_i} \epsilon_{q, \leftarrow} \hat{\mathcal{R}}^\dag     ) \Sigma_q^\dag + h.c.   \right]
\end{equation}
These equations can be used to simulate any combination of forward and backward polarizations.

For the master equation, we use the six collapse operators corresponding to an emission of a photon with polarization $q$ in either the forward or backward direction:
\begin{equation}
 \hat{C}_q =  \sqrt{\Gamma} \, {\Sigma}_q,    \hat{\mathcal{R}},  \quad   \sqrt{\Gamma}  \, \Sigma_q   \hat{\mathcal{R}}^\dag
\end{equation}

Besides the cross-coupling terms which are described later, this concludes the form of the master equation used in the simulation.  The main approximations are 1) 1D motional states  2) cross-coupling scattering and shifts are small 3) photons can only be emitted in the forward and backward direction, and 4) only a pair of counter-propagating beams. 

Now we define a specific polarization for the two beams.  The most conventional polarizations configurations are lin$\perp$lin and $\sigma_+$-$\sigma_-$.  The lin$\perp$lin configuration corresponds to a polarization that changes between circular and linear, and therefore has a strong vector shift. The $\sigma_+$-$\sigma_-$ configuration is instead a linear rotating field and therefore only has a tensor shift.  Because the MOT is $\sigma_+$-$\sigma_-$ and we use the same beams, the cooling also uses $\sigma_+$-$\sigma_-$.
%\subsubsection{Atom-field interaction for $\sigma_+$-$\sigma_-$}
 
 For $\sigma_+$-$\sigma_-$, 
the atom-field interaction Hamiltonian for the counter-propagating beams is:
\begin{equation}
    H_{AF}^{\sigma_+-\sigma_-} = \sum_i \Omega_i  \left[ ( \hat{\mathcal{R}} \Sigma_{1}^\dag + e^{i \phi_i} \hat{\mathcal{R}}^\dag \Sigma_{-1}^\dag ) + h.c.   \right]
\end{equation}
We assume that the atom is situated around a position $z=0$, where it experiences linear polarization determined by the retro phases $\phi_1$ and $\phi_2$ for the two different frequencies.
Since the cooling and repumper lights have different frequencies, their phase accumulation from the atom to the mirror and back differs by $\phi_{1/2} - \phi_{3/2} \approx 0.3 (2 \pi )$. All three retro mirrors are positioned about 10 inches away from the atoms. This phase difference results in the linearly polarized light seen by the atom at the two frequencies having different angles.  

When the phase difference is zero, the two polarizations are linear and aligned. However, in our experiment and simulation, this phase seems to be of little importance, likely because the repumper has 10 times less power than the cooling beam and does not contribute significantly to the AC Stark shifts.

 Performing the LD expansion of this atom-field Hamiltonian gives (for a single frequency)
 \begin{equation}
     H_{AF}^{\sigma_+-\sigma_-} \approx \Omega_i \left[  \left( \Sigma_{1}^\dag  + e^{i \phi_1} \Sigma_{-1}^\dag \right) + i k  \hat{z} \left( \Sigma_{1}^\dag  - e^{i \phi_1} \Sigma_{-1}^\dag   \right) \right]
 \end{equation}
In the LD regime, the zeroth order term does not couple different motional states $|n \rangle$, but does couple the levels within a single motional state and results in the dressed states shown in the cartoons of Fig.~4 in the main paper.  The first order term results in coupling between different motional states. 
Interestingly, it is this sign difference in the zeroth order and first order term that allows the first order coupling term to mix bright and dark states of different $|n\rangle$.  Without that sign difference, the bright and dark states of neighboring $n$ would not be coupled. 

%\subsubsection{Cross-coupling}
The steady-state solver is unable to handle time-dependent interactions. Given that the cross-coupling is off-resonant, we assume that the scattering rate is small compared to that of the main beams. To address this, we derive additional effective cross-coupling collapse operators as follows:
\begin{equation}
    \hat{C}_i = \sqrt{\Gamma \frac{ \Omega_i^2}{4 \Delta_i^2}}  \left[  \Sigma_{1,i}^\dag \hat{\mathcal{R}} + e^{i \phi_i} \Sigma_{-1, i}^\dag \hat{\mathcal{R}}^\dag  \right]  
\end{equation}
where $\Delta_i$ represents the detuning of the cross-coupling beams to the $F'=3/2$ excited state. These collapse operators are derived by first moving to a rotating frame of the cross-coupling beams, then transitioning to the dressed frame of the time-independent interaction, considering only first-order terms in saturation, and finally returning to the rotating frame of the main beams.

The cross-couplings introduce additional AC Stark shifts, which might affect the $F=1/2$ and $F=3/2$ states. However, we find this impact to be minimal and thus ignore it. The primary effect is the introduction of new collapse operators that correspond to decay from the ground to excited state. These new collapse operators serve two purposes: 1) they induce recoil heating at the rate of off-resonant scattering of the cross-coupling beam, and 2) they cause mixing between the dark and bright states.

 \subsubsection{Simulations}
 In the simulation, we simplify the trap by representing it as a one-dimensional model, rather than its actual three-dimensional configuration. Although the real experiment employs three beams with corresponding retro-reflections, our simulations are limited to one dimension. The simulation would more accurately model pulsing the three pairs of beams individually. Using all three beams simultaneously might lead to interference and alter the results. However, experimental evidence suggests that cooling remains robust even with minor beam adjustments, indicating that interference is likely not a significant factor. Future research will explore the differences between simultaneous and pulse-based multi-axis cooling.

\begin{figure}[h]
    \centering
    \includegraphics[width=0.4\textwidth]{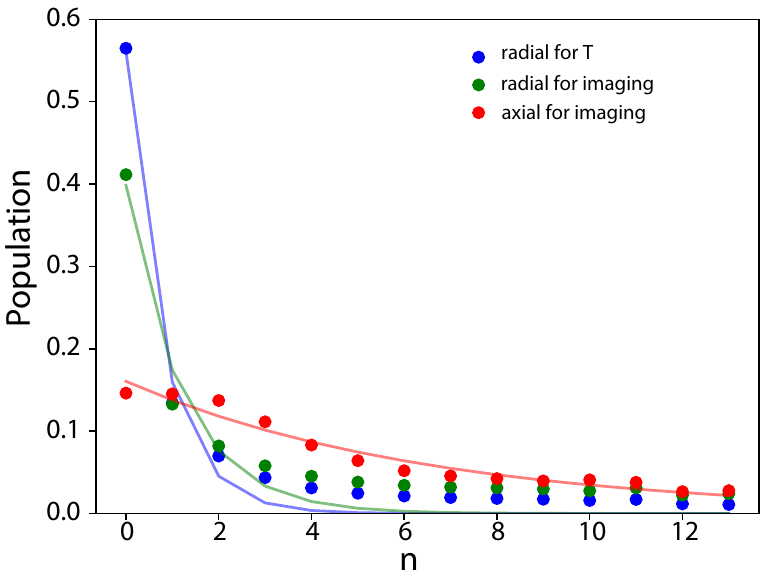}
    \caption{\label{fig:population}    Examples of distributions of population and fits to a Boltzmann distribution to extract temperature.  (blue) Radial population for simulation optimized for radial temperature. (green) radial temperature optimized to minimize high $n$.  (red) axial optimized to minimize high $n$.        }
\end{figure}

 We use QuTip's steady-state solver, employing an iterative approach with up to 14 harmonic states. This solver generates the steady-state density matrix, from which we determine the motional state distribution and excited state population. By fitting the motional state distribution to a Boltzmann distribution $e^{- \beta \hbar \omega n}$, we can extract the temperature $\beta = 1/k_B T $. Fig.~\ref{fig:population} shows the population distributions and Boltzmann fits for several optimized simulations.  The blue curve is when radial temperature is minimized, and gives $T = 0.8 \, \hbar \omega /k_B$ = 27 $ \mu$K, or 57 $\%$ ground state.  The green curve is when radial is instead optimized to minimize the high population.   The red curve is the population when axial is optimized for temperature, and give $T = 3.7 \hbar \omega_\text{axial} / k_BT = $ 27 $\mu$K, or 7$\%$ ground state.   In the 2d simulations of the main of the main text corresponding to the experimental parameters, we get 41 $\mu$K for radial and 82 $\mu$K for axial.  
 
% For the 2D simulations shown in Fig.~4 of the main text, we get 
%  from 2d simulations in paper,  $T = 1.2 (hw/kBT)^-1$,  or  41 uK for radial,  $<n> = 3.0$
%  for axial 12,  or  82 uK.  $<n> = 5.0$
%  0.006 excited for both.  
% $(2*(<nradial> +1/2)*34uK + (<naxial>+1/2)*34/5 uK) = 3 kBT$,  or T = 92 uK. 

%
The radial trapping frequency was measured to be 700 kHz, corresponding to a motional quanta of 33.6 $\mu$ K.  The axial trapping frequency was not directly observed, but is typically 5 -6 times smaller in similar experiments.  From the simulations, we can either extract average 1D energy $\langle  \hbar \omega ( n+1/2) $ or fit the population to a Boltzmann distribution to extract temperature.  We find the population distributions typically follow a Boltzmann distribution and that it converges for less harmonic states.  That temperature is also more accurate for comparing to the temperature extracted from release and recapture.  The temperature extract from both methods typically did not differ much.

Importantly, we see that the axial and radial cooling have very different conditions. Experimentally, we have optimized the combined temperature, but the simulation suggests we could cool further by pulsing the radial and axial cooling separately with different powers.  Indeed in simulation we find the temperature can be reduced by more than a factor of two by optimizing the radial and axial separately. 

The Lamb-Dicke (LD) regime approximation is not employed; however, the values in the $\mathcal{R}$ matrix are discarded if they fall below $5 \times 10^{-3}$ in order to speed up the simulations.  The simulations were executed for all D1 lines for up to 10 harmonic states directly and 20 iteratively using the BICGSTAB algorithm with an iLu preconditioner within QuTip using the Purdue Negishi computing cluster. 

\subsubsection{D1 Saturation Intensity}
In the main text we refer to the D1 saturation intensity.  As there seems to be confusion in literature, we here derive the D1 saturation intensity.  The reduced matrix element convention adopted is given by
\begin{equation}
    \tau^{-1}  =  \frac{\omega_0^3 }{3 \pi \epsilon_0 \hbar c^3} \frac{2 J+1}{2J'+1}  |\langle  J || \mathbf{d} || J' \rangle |^2,
\end{equation}
with $\tau = 27.102$ ns. The reduced matrix element is:
\begin{equation}
    \langle J=1/2 || \mathbf{d} || J'=1/2\rangle = 1.989 \times 10^{-29}  \, \text{C} \cdot \text{m}.
\end{equation}
The saturation intensity of a transition is defined by
\begin{equation}
    \frac{I}{I_\text{sat}} = \frac{  c \epsilon_0 \Gamma^2  \hbar^2  }{4 \, | \langle F, m |\mathbf{d}| F', m'\rangle  |^2}.
\end{equation}
By convention, the saturation intensity for D1 or D2 is cited as the smallest saturation intensity for any two transitions with any polarization. The largest dipole moment of any two transitions in D1 is for $|F=1/2, m=1/2 \rangle $ to $| F'=3/2, m'=3/2 \rangle $, which yields a saturation intensity of 
\begin{equation}
 I_\text{sat}^\text{D1} = 5.72 \,\, \text{mW}/\text{cm}^2.
\end{equation}
We note the disagreement with Ref.~\cite{gehm2003}, where the largest dipole moment was not used. The D1 saturation intensity differs approximately by $\sqrt{2}$ from the more commonly cited D2 saturation intensity, $I_\text{sat}^\text{D2} = 2.54 \,\, \text{mW}/\text{cm}^2$.

\bibliography{bibliography}